\documentclass[preprint,prc,tightenlines,showpacs,superscriptaddress,nofootinbib]{revtex4-1}
\usepackage{amsfonts}
\usepackage{amssymb}
\usepackage{amsmath}
\usepackage{graphicx,color}
\usepackage{dcolumn}
\usepackage{epsfig}
\usepackage{bm}
\usepackage{bbm}
\usepackage{ulem}
\usepackage{hyperref}
\usepackage{lineno}

\usepackage{color}

\newcommand{\ket}[1]{|{#1}\rangle}
\newcommand{\bra}[1]{\langle{#1}|}
\newcommand{\inp}[2]{\langle{#1}|{#2}\rangle}

\def\gtap{\ \raise.3ex\hbox{$>$\kern-.75em\lower1ex\hbox{$\sim$}}\ }
\def\ltap{\ \raise.3ex\hbox{$<$\kern-.75em\lower1ex\hbox{$\sim$}}\ }
\begin{document}

\title{Neutron-neutron scattering length from\\
 $\pi^+$ photoproduction on the deuteron}
\author{Satoshi X. Nakamura}
\affiliation{
State Key Laboratory of Particle Detection and Electronics,
University of Science and Technology of China, Hefei 230036, China}
\affiliation{
Department of Modern Physics,
University of Science and Technology of China, Hefei 230026, 
China~\email{satoshi@ustc.edu.cn}
}

\author{Takatsugu~Ishikawa}
\affiliation{Research Center for Electron Photon Science (ELPH), Tohoku University, Sendai 982-0826, Japan}

\author{Toru Sato}
\affiliation{Research Center for Nuclear Physics (RCNP), Osaka University, Ibaraki 567-0047, Japan}

\begin{abstract}
We discuss the possibility of extracting the neutron-neutron scattering length $a_{nn}$
and effective range $r_{nn}$ from 
cross-section
 data ($d^2\sigma/dM_{nn}/d\Omega_\pi$),
as a function of the
 $nn$ invariant mass $M_{nn}$, 
for $\pi^+$ photoproduction on the deuteron ($\gamma d\to \pi^+nn$).
The analysis is based on a $\gamma d\to \pi^+nn$ reaction model
in which realistic 
elementary amplitudes for $\gamma p\to \pi^+n$, $NN\to NN$, 
and $\pi N\to \pi N$ are incorporated.
We demonstrate that the $M_{nn}$ dependence (line shape)
of a ratio $R_{\rm th}$,
$d^2\sigma/dM_{nn}/d\Omega_\pi$ normalized by
$d\sigma/d\Omega_\pi$ for $\gamma p\to\pi^+ n$ and 
the nucleon momentum distribution inside the deuteron,
at the kinematics with $\theta_\pi=0^\circ$ and $E_\gamma\sim 250$~MeV is particularly
useful for extracting $a_{nn}$ and $r_{nn}$ from 
the corresponding $R_{\rm exp}$ data.
We found that $R_{\rm exp}$ with 2\% error,
 resolved into an $M_{nn}$ bin width of 0.04 MeV (corresponding to a
 $p_\pi$ bin width of 0.05 MeV$/c$), can determine the $a_{nn}$ and $r_{nn}$
 with uncertainties of $\pm 0.21$ and $\pm 0.06$ fm, respectively,
if $a_{nn} = -18.9$~fm and $r_{nn} = 2.75$~fm.
The requirement of such narrow bin widths indicates that
the momenta of the incident photon and emitted $\pi^+$
must be measured at high resolutions.
This can be achieved by
 utilizing virtual photons of very low $Q^2$
from electron scattering at the 
Mainz Microtron
facility.
The method proposed herein for determining the $a_{nn}$ and $r_{nn}$ 
from $\gamma d\to \pi^+ nn$
has a great experimental 
advantage over the previous method of
utilizing $\pi^- d\to\gamma nn$ for not requiring the formidable task of controlling 
the neutron detection efficiency and its uncertainty.
\end{abstract}

\maketitle

\section{Introduction}

Charge symmetry (CS) is an important concept used to describe many facets of
nuclear physics~\cite{cs_review,miller2}.
It leads to a consequence that observables hardly change
when all protons and neutrons in a nuclear system are replaced by 
neutrons and protons, respectively.
For example, the excited states of mirror nuclei have
identical energy levels and spin-parity assignments.
The CS can be more generally defined by the invariance under the rotation
 by $180^\circ$ about the $y$-axis in the isospin space.
This rotation corresponds to 
 the interchange of $u$ and $d$ quarks at the quark level and
the interchange of protons and neutrons at the hadron level.

Within the Standard Model,
the CS is broken 
due to the differences among 
$u$ and $d$ quark masses and electromagnetic (EM) effects.
These elementary effects appear in hadron phenomenology in various ways,
e.g., the neutron ($n$) and proton ($p$) mass difference of 1.3~MeV
and the small charge-dependent component of nuclear force that
breaks the CS at the order of a few percentages. 
In terms of 
the hadronic degrees of freedom,
the CS breaking (CSB) of nuclear force can be described
by 
a mixing of the neutral rho meson ($\rho^0$)
and omega meson ($\omega$) 
in
one boson exchange mechanism~\cite{miller2}
and the $n$-$p$ mass difference.
This CSB force could explain the following experimental observations:
the 0.7-MeV difference in the binding energies
between ${}^3$H and ${}^3$He (mirror nuclei);
the difference in the analyzing powers
$A_n(\theta_n)\neq A_p(\theta_p)$
at the same angle $\theta_n=\theta_p$ for $np$ scattering~\cite{abegg};
the forward-backward asymmetry
$d\sigma/d\Omega_{\pi}(\theta)\neq d\sigma/d\Omega_\pi(\pi-\theta)$
in the deuteron ($d$) formation reaction emitting a neutral pion ($np\to d\pi^0$)~\cite{csb_np_dpi0}.
A large difference in the excitation energies
between the $A = 4$ mirror hypernuclei, i.e., $^4_\Lambda$H and $^4_\Lambda$He,
has been recently reported~\cite{csb_hyper_jparc,csb_hyper_a1}.
It is possible that CSB occurs
more strongly in hypernuclei than in ordinary nuclei.
A complete understanding of CSB still remains an open issue in nuclear 
physics.
To reveal the cause of the CSB observed in few-baryon systems,
it is of critical importance to experimentally investigate
the differences between low-energy elementary $nn$ and $pp$ scatterings
as well as between $\Lambda n$ and $\Lambda p$ scatterings.

Low-energy $NN$ scattering is characterized 
by the scattering length $a$ and effective range $r$
through an effective-range expansion of the $S$-wave phase shift $\delta(p)$ as follows:
\begin{equation}
p \cot\delta(p) = -\frac{1}{a} + \frac{1}{2} \, r\, p^2 + O(p^4),
\label{eq:ERE}
\end{equation}
where $p$ denotes the momentum of the nucleon ($N$)
in the $NN$ center-of-mass (CM) frame.
Note that a positive or negative $a$ value indicates
a repulsion or attraction, respectively, in this definition.
The experimentally obtained $a$ and $r$ parameters of the spin-singlet ($^1\!S_0$) states are: 
\begin{equation}
\left\{
\begin{array}{lll}
a_{nn} = -18.9\pm 0.4 {\rm \ fm}, & r_{nn} = 2.75 \pm 0.11 {\rm \ fm} &{\rm for\ } nn,\\
a_{np} = -23.74\pm 0.02 {\rm \ fm}, & r_{np} = 2.77 \pm 0.05 {\rm \ fm} &{\rm for\ } np, {\rm \ and}\\
a_{pp} = -17.3\pm 0.4 {\rm \ fm}, & r_{pp} = 2.85 \pm 0.04 {\rm \ fm} &{\rm for\ } pp,\\
\end{array}
\right.
\label{eq:ann}
\end{equation}
where the EM effects have already been corrected.
The scattering length $a_{np}$ in the $np$ system is significantly different 
from the other two, i.e., 
$a_{nn}$ and $a_{pp}$, 
suggesting 
charge independence breaking.
The CSB is significant at the 1.6-fm difference between $a_{nn}$ and $a_{pp}$.
Moreover, while the error of $a_{nn}$ predominantly stems from the statistical uncertainty in the experiments,
that of $a_{pp}$ originates from the systematic uncertainty of removing the EM effects.

Thus far, several different $a_{nn}$ values ranging from $-19$ to $-16$ fm
have been reported (see Ref.~\cite{gardestig_review} for a recent
review),
but no consensus has been reached on which value is correct.
One experimental difficulty is that conducting
an $nn$ scattering experiment,
through which the $a_{nn}$ value can be determined directly, is nearly impossible
because a realistic free neutron target does not exist.
Therefore, the primary experimental results for $a_{nn}$ are from two
different types of experiments utilizing
the final-state $nn$ interaction (indirect determination):
(i) the three-body breakup reaction of $n d\to nnp$;
(ii) the radiative capture of a stopped negative pion on the deuteron
($\pi^-d \to nn\gamma$).

For type (i) experiments,
excluding those before 1973, 
$a_{nn}$ values are extracted from the data with
the exact solution of the Faddeev equation for $n d\to nnp$~\cite{faddeev}.
Significantly different $a_{nn}$ values have been obtained from the same reaction:
\begin{equation}
\left\{
\begin{array}{l@{}l@{}l}
a_{nn} = -16.1\pm 0.4 {\rm \ fm\ } & (E_{n} = 25.3{\rm \ MeV}, &{\ }np  {\rm \ detected}~\text{\cite{huhn}}),\\
a_{nn} = -18.7\pm 0.7 {\rm \ fm\ } & (E_{n} = 13.0{\rm \ MeV}, &{\ }nnp {\rm \ detected}~\text{\cite{gonz}}), {\rm\ and}\\
a_{nn} = -16.5\pm 0.9 {\rm \ fm\ } & (E_{n} = 17.4{\rm \ MeV}, &{\ }p   {\rm \ detected}~\text{\cite{wits}}).\\
\end{array}
\right.
\end{equation}
The primary concern with these results is the possibility of large 
three-body force effects.
As the details of these effects are not yet well-established,
the systematic uncertainty associated with them could be underestimated.

The type (ii) experiments of $\pi^-d \to nn\gamma$ are
considered as
a more reliable method to determine the $a_{nn}$ value
because only $nn$ scattering without three-body force effects
occurs in the final state.
The obtained experimental value is $a_{nn} = -18.9\pm 0.4$~fm
after including the correction of 
$\Delta a_{nn} \sim -0.3$~fm 
from the magnetic-moment interaction between $nn$~\cite{chen}.
The difficulty of experimentally studying $\pi^- d \to nn \gamma$ 
lies in detecting
low-energy neutrons. 
The efficiency of detecting 
the
neutrons 
depends on their kinetic energies
and is sensitive to the detector threshold
measured in the electron-equivalent energy.
In Refs.~\cite{chen,howell}, the detector efficiencies were checked at
neutron kinetic
energies from 5 to 13 MeV using energy-tagged neutrons produced in the
${}^2{\rm H}(d,n){}^3{\rm He}$ reaction. 
However,
a direct measurement has not yet been performed for detector
efficiencies around $\sim$2.4~MeV, at which neutrons affected by the final-state
interaction are expected to appear from $\pi^-d\to nn\gamma$.
Regarding theory, 
a series of works have been conducted 
based on phenomenological~\cite{GGS} and 
dispersion-relation approaches~\cite{Teramond}.
A more recent 
 work has also been 
done
based on the 
 chiral effective field theory~\cite{anders}.
Due to the dominance of the Kroll-Ruderman term,
the pion photoproduction amplitude is rather well-controlled.
It has been reported that 
the primary theoretical uncertainty stems from the off-shell behavior of the
$nn$ rescattering amplitude.

Another possible method
of determining the $a_{nn}$ value is to utilize 
the final-state $nn$
interaction in the $\gamma d \to \pi^+nn$ reaction.
This possibility was 
pointed out
by Lensky {\it et al.}~\cite{prev},
who studied the reaction with the chiral perturbation theory.
However,
their calculation is limited to the energy region close to the pion-production
 threshold
(photon energy up to 20~MeV above the threshold, 
 corresponding to an emitted pion momentum less than 80 MeV$/c$).
 It is difficult to experimentally detect such a low-momentum $\pi^+$
 before its decay.
Alternatively, let us consider the reaction at
 an incident photon energy of $E_\gamma = 200$--300~MeV.
This energy region
is between the pion production threshold ($E_\gamma\sim 150$ MeV)
and the excitation energy of the delta baryon ($\Delta(1232)\,P_{33}$)
($E_\gamma\sim 340$ MeV),
 assuming that the quasi-free $\gamma p \to \pi^+ n$ reaction occurs on
 the initial nucleon at rest inside the deuteron. 
Here, we choose to detect the
 $\pi^+$s
emitted at $\theta_\pi\simeq 0^\circ$ from the photon direction,
and only those near the maximum momentum are of interest. 
In this particular 
kinematics,
the relative momentum of $nn$ is
low, and $nn$ are expected to strongly interact with each other.
Additionally, this would efficiently prevent
a pion created in $\gamma N \to \pi N$ 
from rescattering on the spectator nucleon
because the $\pi N$ interaction is weak at low energies 
and/or
the spectator nucleon 
is required to
have a large momentum, which is largely suppressed in the deuteron.
This seems to be an ideal condition with which to study low-energy neutron-neutron
scattering, thereby determining the $a_{nn}$
as well as $r_{nn}$ values.

For extracting $a_{nn}$ from $\pi^- d\to\gamma nn$
or $\gamma d\to\pi^+ nn$,
the experimental challenge is to obtain
the $nn$ invariant mass ($M_{nn}$) distribution
at $M_{nn}\sim 2 m_{n}$
with high statistics and high resolution.
In this respect, 
we find that
$\gamma d\to\pi^+ nn$ 
is more advantageous than $\pi^- d\to\gamma nn$
because we can avoid neutron detection,
the efficiency of which could significantly increase
the systematic uncertainty.
We only need to detect $\pi^+$s
with 
momenta
of 120--250 MeV/$c$
once the incident photon energies have been determined 
to a sufficient precision.
Thus, it seems valuable 
to analyze
$\gamma d\to\pi^+ nn$ data 
and extract the $a_{nn}$ value, which is both
independent of and alternative to those from previous methods using
$\pi^- d\to\gamma nn$ and $nd$ scattering data.
To extract the $a_{nn}$ and $r_{nn}$
from the $\gamma d\to\pi^+ nn$ data in a controlled manner,
we must estimate
possible theoretical uncertainties.
Therefore, in this work,
we use a theoretical model for the $\gamma d\to \pi^+ nn$ reaction
and examine the reaction at the
particular kinematical conditions of 
$E_\gamma = 200$--300 MeV, 
$\theta_\pi\sim 0^\circ$, and $M_{nn}\sim 2 m_{n}$.
 Through a theoretical analysis of 
 $\gamma d \to \pi^+nn$,
 we find that the kinematics
and shape of the $M_{nn}$ distribution
are indeed suitable 
for studying low-energy $nn$ scattering.
We also assess possible theoretical uncertainties 
of the $M_{nn}$ distribution needed
when extracting the $a_{nn}$ and $r_{nn}$ values
 from the corresponding data.
Finally, we conduct a Monte Carlo simulation to extract 
the $a_{nn}$ and $r_{nn}$ from the data, 
prepared with different precisions and
$M_{nn}$ bin widths,
and estimate their uncertainties.
This analysis leads to the proposal of 
an alternative and more reliable method of extracting the $a_{nn}$ 
and $r_{nn}$.

The rest of this paper is organized as follows:
In Sec.~\ref{sec:formalism}, we discuss the theoretical formalism used to
study $\gamma d \to \pi^+nn$.
The numerical results are presented and discussed in Sec.~\ref{sec:result}.
Sec.~\ref{sec:virtual} is devoted to a discussion on 
the experimental strategy of 
measuring $\gamma d \to \pi^+nn$ with high resolution
using virtual photons of a 
low $Q^2$ ($<0.01$~GeV$^2$).
Finally, a summary follows in Sec.~\ref{sec:summary}.

\section{Formalism}
\label{sec:formalism}

  \subsection{The $\gamma d\to \pi^+ nn$ reaction model based on the dynamical
  coupled-channels model}
\label{sec:2.1}
  Our starting point to 
develop
  a $\gamma d\to \pi^+ nn$ reaction model is
  the elementary amplitudes for the
  $\gamma N\to \pi N$ and $\pi N\to \pi N$ processes.
  Here, we employ
the elementary amplitudes generated by a dynamical
  coupled-channels (DCC) model~\cite{knls13,knls16}.
  The DCC model includes meson-baryon channels
  relevant to the nucleon resonance and $\Delta$ resonance 
  (generically referred to as $N^*$)
  region, such as 
  $\pi N, \eta N, K\Lambda, K\Sigma$, and also
  $\pi\Delta, \sigma N, \rho N$ which couple to $\pi\pi N$.
  The $\gamma^{(*)} N$ channel is also considered perturbatively.
  The meson-baryon interaction potentials comprise
  meson-exchange non-resonant and (bare) $N^*$-excitation resonant
  mechanisms.
  The gauge invariance is satisfied at the tree level.
  Upon solving the coupled-channel Lippmann-Schwinger equation, with off-shell
  effects fully considered, the unitary DCC amplitudes are obtained.
  The DCC model was developed through a comprehensive analysis of the
  $\pi N, \gamma N\to\pi N, \eta N, K\Lambda$, and $K\Sigma$ data in the
  CM energy ($W$) region from the channel thresholds to $W\ltap 2.1$~GeV.
  The model provides a reasonable description of the data included in the
  fits, and the properties of all the well-established nucleon
  resonances have been
  extracted from the obtained amplitudes.
  The DCC model was extended to a finite $Q^2$ region via analyzing
  electron-induced data~\cite{dcc_nu} as well as to neutrino-induced
  reactions via developing the axial current amplitudes~\cite{dcc_nu,nu-review,nu-comp}.

\begin{figure}[t]
\begin{center}
\includegraphics[width=1\textwidth]{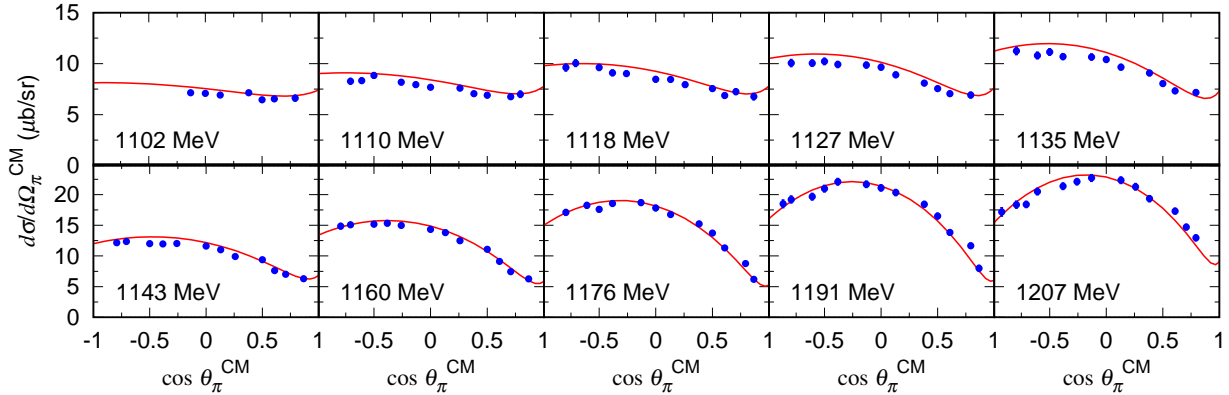}
\end{center}
 \caption{
 Differential cross sections of $\gamma p\to\pi^+n$ from the DCC model.
 The CM energy $W$ is indicated in each panel. 
The $W = 1121$, 1162, and 1201~MeV energies correspond to the incident
 photon energies of $E_\gamma \simeq 200$, 250, and 300~MeV, respectively. 
The data are taken from Ref.~\cite{gp-data}.
The errors shown are statistical only.
 }
 \label{fig:xs_gp}
  \end{figure}
  The $\gamma p\to \pi^+ n$ elementary amplitude is of primary importance
  when developing a $\gamma d\to\pi^+ nn$ model.
Therefore, it is reassuring to see in Fig.~\ref{fig:xs_gp} that
  the DCC model well describes the
  $\gamma p\to \pi^+ n$ cross-section data in the energy region relevant
  to this work.
However, data are unavailable for the forward direction
($\cos\theta_\pi^{\rm \,CM} = 1$), 
which is particularly important for our purpose. 
Moreover, the data shown in Fig.~\ref{fig:xs_gp} have additional
systematic errors.
These facts could raise concern as to whether we 
can develop
a $\gamma d\to\pi^+ nn$ model 
that reaches the precision required to extract the $a_{nn}$ from data.
As will be discussed later, when extracting the $a_{nn}$,
we use a method 
that largely cancels out the normalization uncertainty of the elementary 
$\gamma p\to \pi^+ n$ amplitudes.

  With the DCC model as the starting point, it is straightforward to
  extend the model
  to a $\pi NN$ system following the well-established multiple
  scattering theory~\cite{kmt}.
  In this work, we consider the impulse and first-order
  rescattering terms, as depicted in Fig.~\ref{fig:diag},
wherein higher order rescattering terms are truncated.
  This setup, including up to the first-order rescattering, has been
  used in 
previous
works~\cite{arenhover,fix,lev06,sch10,wsl15,tara-1,dcc-deu4}
 and shown to provide a reasonable description of $\gamma d\to \pi NN$ data
with significant rescattering effects. 
  As we will see, this truncation is a good approximation
  for the particular kinematics considered in this work.

  Similar DCC-based deuteron reaction models have
   been successfully applied to solve
   several problems of current interest. 
  We proposed a novel method to determine the $\eta N$ scattering
  length using $\gamma d\to \eta pn$ data at a special kinematics in
  Ref.~\cite{dcc-deu1}.
  In Ref.~\cite{dcc-deu2}, the extraction of neutron-target observables
  from $\gamma d\to \pi NN$ data was examined, and some rescattering effects
observed in the data were elucidated for the first time.
In Ref.~\cite{dcc-deu3}, neutrino-nucleon cross sections
from neutrino-deuteron data were corrected by estimating the rescattering effects, significantly contributing to neutrino oscillation experiments~\cite{nustec}.

  \subsection{Cross-section formula}

\begin{figure}[t]
\begin{center}
\includegraphics[width=.9\textwidth]{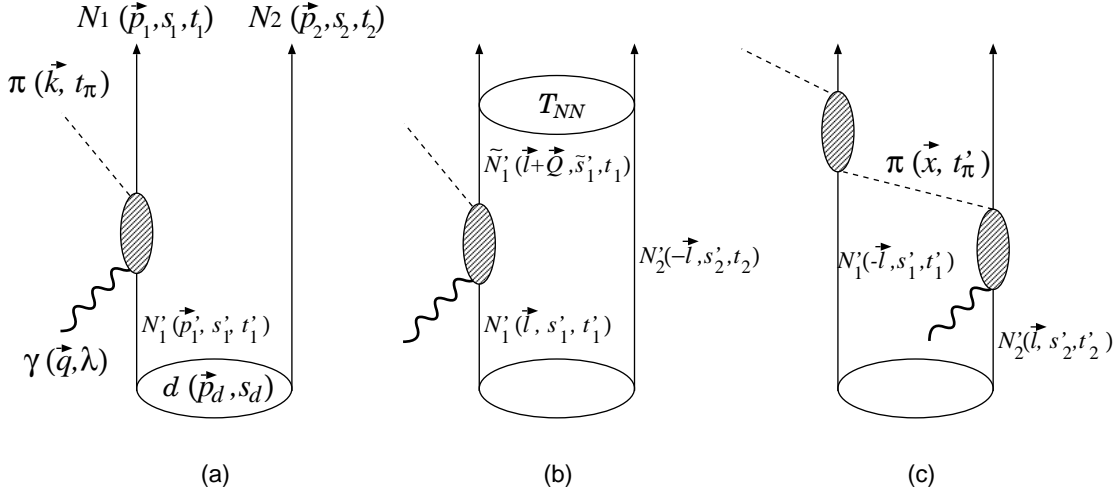}
\end{center}
 \caption{
 The diagrams for $\gamma d\to \pi^+ nn$ considered in this work;
(a) impulse, (b) $NN$ rescattering, and (c) $\pi N$ rescattering mechanisms.
 We introduced $\bm{Q}\equiv\bm{q}-\bm{k}$ and
 $\bm{x}\equiv\bm{q}-\bm{p}_2+\bm{l}$.
 The shaded ellipses
are elementary amplitudes from the DCC model.
 }
\label{fig:diag}
\end{figure}
The unpolarized differential cross-section formula for
$\gamma(\bm{q})+ d(\bm{p}_d)\to \pi^+(\bm{k})+n_1(\bm{p}_1)+n_2(\bm{p}_2)$
in the laboratory frame ($\bm{p}_d = 0$) is given as:
\begin{eqnarray}
\frac{d^{2}\sigma(E_\gamma)}{d\Omega_{\bm{k}}dM_{nn}}
= 
{1\over 12}
\sum_{\lambda,s_d}
\sum_{s_1,s_2}
\frac{(2\pi)^4}{4E_\gamma}\frac{1}{2E_d(\bm{p}_d)}
\int d\Omega_{\bm{p}_{nn}} {p_{nn} k^2 m_n^2\over |k E -
\bm{q}\!\cdot\!\hat{k}\, E_\pi(\bm{k})|}\;
|M(E)|^2 \ ,
\label{eq:dsigp-nn}
\end{eqnarray}
where $\lambda$ is the photon polarization and $s_d$ ($s_i$)
is the $z$-component of the deuteron (neutron~$i$) spin,
with a factor of $1/12$ for averaging the initial spins ($1/2\times 1/3$)
and for the identity of the final two neutrons ($1/2$).
The energy $E_x$
for particle $x$ 
depends on the particle mass ($m_x$) and momentum ($\bm{p}_x$) as 
$E_x = \sqrt{\bm{p}^2_x + m^2_x}$, 
and 
the total energy in the laboratory frame is $E = m_d + E_\gamma$.
The $\bm{p}_{nn}$ denotes the momentum of a neutron in the $n_1 n_2$ CM frame.
We simply wrote the magnitudes of the momenta as
$k\equiv |\bm{k}|$ and $\hat{k}\equiv \bm{k}/|\bm{k}|$. 
The quantity $M(E)$ is the Lorentz-invariant amplitude defined below.

As discussed in subsection \ref{sec:2.1},
we describe the $\gamma d\to\pi^+ nn$ reaction
by considering the impulse [$t_{{\rm imp}}$, Fig.~\ref{fig:diag}(a)],
$NN$ rescattering [$t_{NN}$, Fig.~\ref{fig:diag}(b)],
and $\pi N$ rescattering [$t_{\pi N}$, Fig.~\ref{fig:diag}(c)] mechanisms.
The corresponding amplitudes are written using the kinematical variables defined in
Fig.~\ref{fig:diag} as follows (see Ref.~\cite{dcc-deu4} for a more detailed discussion):
\begin{eqnarray}
t_{{\rm imp}} &=&
\sqrt{2}
\sum_{s_1',t'_1}
\bra{\pi(\bm{k},t_\pi)\, N_1(\bm{p}_1,s_1,t_1)} 
t_{\pi N,\gamma N}(M_{\pi N_1})
\ket{\gamma(\bm{q},\lambda)\, N_1'(-\bm{p}_2 ,s_1',t'_1)}
\nonumber\\
&&\times
\inp{N_1'(-\bm{p}_2,s_1',t'_1)\, N_2 (\bm{p}_2,s_2,t_2)}{\Psi_d(s_d)}
\ ,
\label{eq:amp_imp}
\\
t_{NN} &=&\sqrt{2}
\sum_{s_1',\tilde s_1',s_2',t'_1}
\int d\bm{l}\
\nonumber\\
&&\times
\bra{N_1(\bm{p}_1,s_1,t_1)\, N_2(\bm{p}_2,s_2,t_2)}
t_{NN,NN}(M_{N_1N_2}) 
\ket{\tilde N'_1(\bm{q}-\bm{k}+\bm{l},\tilde s'_1,t_1)\, N'_2(-\bm{l},s'_2,t_2)}
\nonumber\\
&&\times
{\bra{\pi(\bm{k},t_\pi)\, \tilde N_1'(\bm{q}-\bm{k}+\bm{l},\tilde s_1',t_1)} 
t_{\pi N,\gamma N}(W)
\ket{\gamma(\bm{q},\lambda)\, N_1'(\bm{l},s_1',t'_1)}
\over E-E_N(\bm{q}-\bm{k}+\bm{l})-E_N(-\bm{l})-E_\pi(\bm{k})+i\epsilon}
\nonumber\\
&&\times
\inp{N_1'(\bm{l},s_1',t'_1)\, N_2' (-\bm{l},s_2',t_2)}{\Psi_d(s_d)}
\ , {\rm and}
\label{eq:amp_NN}
\\
t_{\pi N} &=& \sqrt{2}
\sum_{s_1',s_2'}
\sum_{t_1',t_2',t'_\pi}
\int d\bm{l}
\bra{\pi(\bm{k},t_\pi)\, N_1(\bm{p}_1,s_1,t_1)}
t_{\pi N,\pi N}(M_{\pi N_1})
\ket{\pi(\bm{q}-\bm{p}_2+\bm{l},t'_\pi)\, N_1'(-\bm{l},s'_1,t_1')}
\nonumber\\
&&\times
{\bra{\pi(\bm{q}-\bm{p}_2+\bm{l},t'_\pi)\, N_2(\bm{p}_2,s_2,t_2)} 
t_{\pi N,\gamma N}(W)
\ket{\gamma(\bm{q},\lambda)\, N_2'(\bm{l},s_2',t_2')}
\over 
E-E_N(\bm{p}_2)-E_N(-\bm{l})-E_{\pi}(\bm{q}-\bm{p}_2+\bm{l})+i\epsilon}
\nonumber\\
&&\times
\inp{N_1'(-\bm{l},s_1',t_1')\, N_2' (\bm{l},s_2',t_2')}{\Psi_d(s_d)}
\ .
\label{eq:amp_MN}
  \end{eqnarray}
The exchange terms are obtained 
from Eqs.~(\ref{eq:amp_imp})--(\ref{eq:amp_MN}) 
by flipping the overall sign 
and interchanging 
all subscripts 1 and 2 for nucleons in the intermediate and final $\pi NN$ states.
In the above expressions,
the deuteron state with spin projection $s_d$ is denoted as $\ket{\Psi_d(s_d)}$;
$\ket{N(\bm{p},s,t)}$ is the nucleon state with momentum $\bm{p}$ and spin and isospin projections 
$s$ and $t$, respectively;
$\ket{\gamma(\bm{q},\lambda)}$ is the photon state with momentum $\bm{q}$ and polarization $\lambda$;
$\ket{\pi(\bm{k},t_\pi)}$ is the pion state with momentum $\bm{k}$ and 
isospin projection $t_\pi$.
The two-body elementary amplitudes depend on 
the $\pi N_1$ and $N_1 N_2$ invariant masses 
given by
\begin{eqnarray}
M_{\pi N_1}&=&\sqrt{[E_\pi(\bm{k})+E_{N}(\bm{p}_1)]^2-(\bm{k}+\bm{p}_1)^2}\ , 
{\rm and}
\\
M_{N_1 N_2}&=&\sqrt{[E_{N}(\bm{p}_1)+E_{N}(\bm{p}_2)]^2-(\bm{p}_1+\bm{p}_2)^2}\ , 
\end{eqnarray}
respectively, and also the $\gamma N$ invariant mass 
calculated according to the spectator approximation:
\begin{eqnarray}
W&=&\sqrt{[E-E_{N}(-\bm{l})]^2-(\bm{l}+\bm{q})^2}\ .
\label{eq:w}
\end{eqnarray}
The two-nucleon energy in the propagator of the $NN$ rescattering amplitude 
in Eq.~(\ref{eq:amp_NN})
is calculated via a non-relativistic approximation:
\begin{eqnarray}
E_N(\bm{q}-\bm{k}+\bm{l})+E_N(-\bm{l}) \simeq
\sqrt{(2m_N)^2 + (\bm{q}-\bm{k})^2} + {(\bm{q}/2-\bm{k}/2+\bm{l})^2\over
m_N} \ .
\end{eqnarray}
Finally, the Lorentz-invariant scattering amplitude [$M(E)$]
used in Eq.~(\ref{eq:dsigp-nn}) is expressed with the amplitudes of
Eqs.~(\ref{eq:amp_imp})--(\ref{eq:amp_MN}) by:
\begin{eqnarray}
M(E) &=&
\sqrt{  { 8  E_\gamma E_d(\bm{p}_d) E_\pi(\bm{k})
E_n(\bm{p}_1)E_n(\bm{p}_2)}\over m_n^2}\nonumber \\
&&\times
\Big(
t_{{\rm imp}}(E) +
t_{NN}(E) +
t_{\pi N}(E) +
\{
{\rm exchange\ terms}
\}
\Big) \ .
\label{eq:amp_decomp}
\end{eqnarray}

Regarding the off-shell elementary amplitudes
used in Eqs.~(\ref{eq:amp_imp})--(\ref{eq:amp_MN}),
we use those generated by the DCC
model for the pion photoproduction amplitudes
($t_{\pi N,\gamma N}$) and pion-nucleon scattering amplitudes ($t_{\pi N,\pi N}$).
Moreover,
we generate the half off-shell $NN$ scattering amplitudes ($t_{NN,NN}$) and the deuteron wave
function ($\Psi_d$)
using high-precision phenomenological $NN$ potentials.

\section{Results and discussion}
\label{sec:result}

\begin{figure}[t]
\begin{center}
\includegraphics[width=1\textwidth]{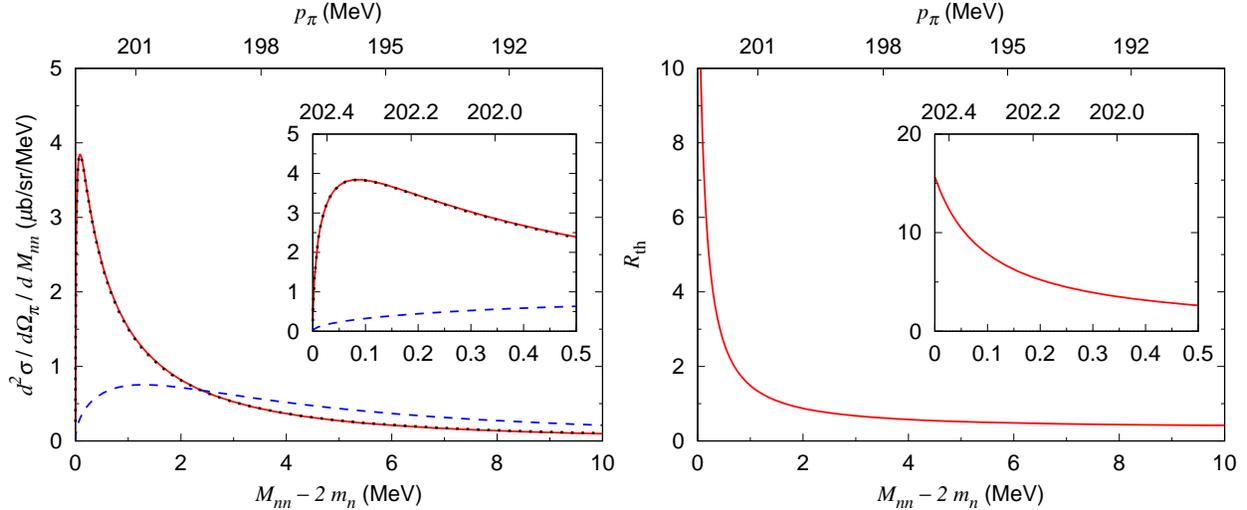}
\end{center}
 \caption{
(Left) The $nn$ invariant mass ($M_{nn}$) distribution
 for $\gamma d\to\pi^+ nn$.
 The incident photon energy is $E_\gamma = 250$~MeV, and
 the pion emission angle is fixed at $\theta_\pi = 0^\circ$.
The blue dashed, black dotted, and red solid curves are calculated with
 the impulse, impulse + $NN$ rescattering, and 
 impulse + $NN + \pi N$ rescattering mechanisms, respectively.
 The emitted pion momentum is indicated on the upper horizontal axis.
The inset shows the small $M_{nn}$ region.
 (Right) Ratio $R_{\rm th}$ defined in Eq.~(\ref{eq:rth})
 for the same condition as the left panel.
 }
\label{fig:xs}
\end{figure}
In Fig.~\ref{fig:xs}(left), we show
the neutron-neutron invariant mass ($M_{nn}$) distribution
($d^2\sigma/dM_{nn}/d\Omega_\pi$ as a function of $M_{nn}$)
for $\gamma d\to\pi^+ nn$ at $E_\gamma = 250$~MeV.
Here, we use the CD-Bonn potential~\cite{cdbonn} 
to describe the $NN$ rescattering and the deuteron wave function.
The pion emission angle is fixed at $\theta_\pi = 0^\circ$.
We focus on a small $M_{nn}$ region, 
as we are interested in the neutron-neutron scattering there.
The impulse contribution
has a quasi-free peak at $M_{nn}-2m_n\sim 1.3$~MeV.
In this particular kinematics, 
the $NN$ rescattering contribution is large and creates a sharp
peak at $M_{nn}\sim 2m_n$.
This is due to
the strong $NN$ interaction in the 
${}^1\!S_0$ wave at low energies.
On the other hand,
the $\pi N$ rescattering contribution hardly changes
the cross sections.
This indicates that the pion is well separated from the $nn$ system in
this kinematics;
thus, the multiple scattering effect beyond the first-order rescattering
is negligible.
As such, this is a favorable kinematics for
our model considering the diagrams in Fig.~\ref{fig:diag}.

Next, we introduce a ratio $R_{\rm th}$, defined by:
\begin{eqnarray}
 R_{\rm th} =
  \left[
\frac{d^2\sigma(E_\gamma)} {d\Omega_{\bm{k}}dM_{nn}}
  \right]
\bigg/
\left[
\frac{d^2\sigma_{\rm conv}(E_\gamma)}{d\Omega_{\bm{k}}dM_{nn}}
\right]
 \ ,
\label{eq:rth} 
\end{eqnarray}
where the numerator is given in
Eq.~(\ref{eq:dsigp-nn}) including
the impulse + $NN + \pi N$ rescattering mechanism.
The denominator is defined by:
\begin{eqnarray}
\frac{d^2\sigma_{\rm conv}(E_\gamma)}{d\Omega_{\bm{k}}dM_{nn}}
 =
\frac{d\sigma_{\gamma p\to \pi^+n}(E_\gamma)}{d\Omega_{\bm{k}}}
\int d^3p_s {m_n\over E_n(p_s)} |\Psi_d(p_s)|^2
\delta(M_{nn}-{\cal M}_{nn}(\bm{p}_s,E_\gamma))
 \ ,
\label{eq:denom}
\end{eqnarray}
with
\begin{eqnarray}
 {\cal M}_{nn}(\bm{p}_s,E_\gamma) =
  \sqrt{(E_n(\bm{q}-\bm{k}-\bm{p}_s)+E_n(\bm{p}_s))^2
  -(\bm{q}-\bm{k})^2  
}
 \ ,
\label{eq:calm}
\end{eqnarray}
and $d\sigma_{\gamma p\to \pi^+n}/d\Omega_{\bm{k}}$
is the differential cross section for $\gamma p\to \pi^+n$
in the laboratory frame.
Eq.~(\ref{eq:denom}) could be derived from
Eq.~(\ref{eq:dsigp-nn}) via the following prescriptions:
(i) consider only the impulse mechanism;
(ii) use the $\gamma p\to \pi^+n$ amplitudes, assuming that the initial
proton is at rest and free (without binding energy) inside the deuteron;
(iii) ignore the interference with the exchange term;
(iv) ignore the deuteron $D$-wave contribution.
The deviation of 
$R_{\rm th}$ from one is a rough measure of the rescattering effects. 
The practical advantage of using $R_{\rm th}$ 
over the cross section itself is that 
the overall normalization uncertainty
of the $\gamma p\to \pi^+n$ amplitudes
is largely canceled.
This uncertainty is primarily a carry-over from 
that in the $\gamma p\to \pi^+n$ data fitted.
The ratio is also advantageous from an experimental viewpoint
because
the counterpart $R_{\rm exp}$ is given with 
measurable cross sections for proton and deuteron targets
and the deuteron wave function;
the systematic uncertainty
in the overall normalization associated with the number of incident particles,
the target thickness, and the solid angle for pion detection is also
largely canceled in $R_{\rm exp}$.
We present $R_{\rm th}$ in Fig.~\ref{fig:xs}(right)
at the same kinematical setting as the left panel.
Once again, a strong rescattering effect can clearly be observed at
$M_{nn}\sim 2 m_n$.

\begin{figure}[t]
\begin{center}
\includegraphics[width=1\textwidth]{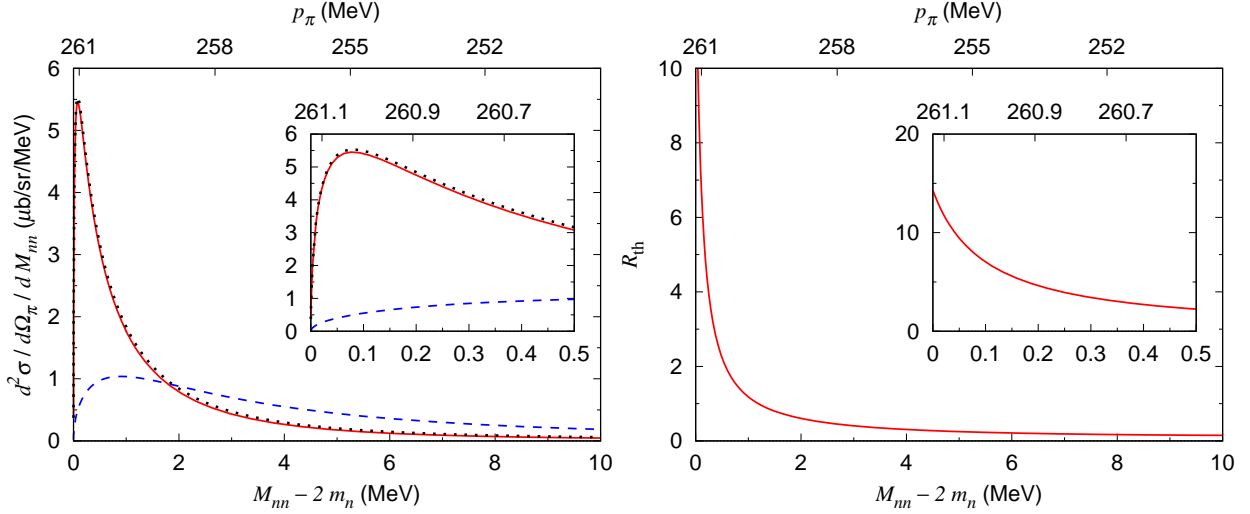}
\end{center}
 \caption{
 Same as in Fig.~\ref{fig:xs} but for the incident photon energy of $E_\gamma = 300$ MeV.
 }
\label{fig:xs-300}
\end{figure}

\begin{figure}[t]
\begin{center}
\includegraphics[width=1\textwidth]{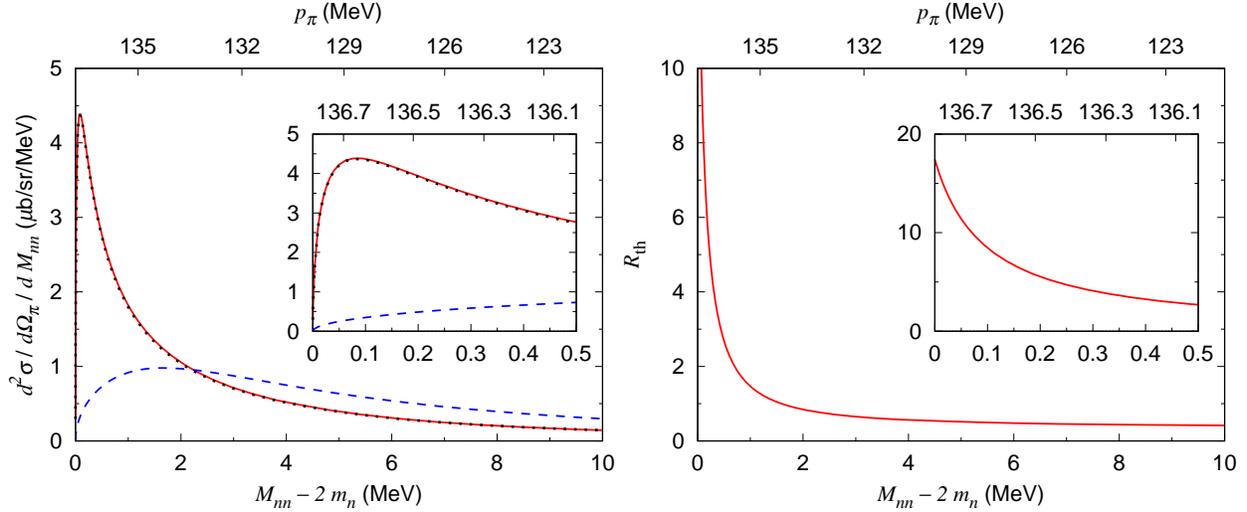}
\end{center}
 \caption{
 Same as in Fig.~\ref{fig:xs} but for the incident photon energy of $E_\gamma = 200$~MeV.
 }
\label{fig:xs-200}
\end{figure}
Next, we examine the results at different $E_\gamma$ values.
In Fig.~\ref{fig:xs-300}, the $d^2\sigma/d\Omega_\pi/dM_{nn}$ and $R_{\rm th}$ are given
for $E_\gamma = 300$~MeV.
While the line shapes are rather similar to those in Fig.~\ref{fig:xs},
the pion rescattering effect is discernible at this energy.
To reduce the theoretical uncertainty, it is best to avoid kinematics
wherein the pion rescattering effect is non-negligible.
The result for $E_\gamma = 200$~MeV is shown in Fig.~\ref{fig:xs-200},
which is qualitatively similar to Fig.~\ref{fig:xs}.
At low photon energies, a contribution from
the elementary $\gamma p\to\pi^+n$ amplitude in the subthreshold
region could be 
non-negligible.
When the loop momentum becomes large, as in 
Figs.~\ref{fig:diag}(b) and (c),
the $W$ for the elementary $\gamma p\to\pi^+n$ amplitude, given by Eq.~(\ref{eq:w}),
falls below the $\pi N$ threshold.
We found this contribution to be
$\sim 2$\% to the $\gamma d\to\pi^+ nn$ cross sections.
As data are absent for testing the subthreshold amplitude,
it is best to use a higher $E_\gamma$
to suppress this contribution.
We observed that the subthreshold contribution was negligible
for $E_\gamma\gtap 250$~MeV.
Thus, 
we choose to study $\gamma d\to\pi^+ nn$ at $E_\gamma = 250$~MeV.

\begin{figure}[b]
\begin{center}
\includegraphics[width=.6\textwidth]{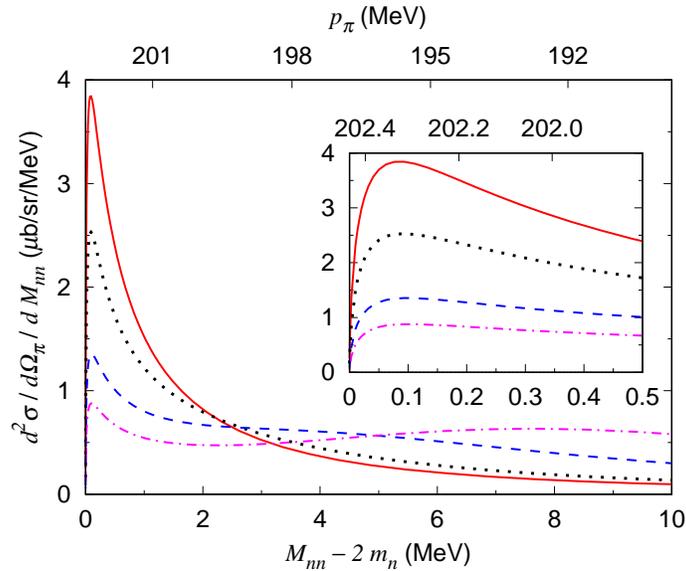}
\end{center}
 \caption{
The $M_{nn}$ distributions for $\gamma d\to\pi^+ nn$ 
 at different pion-emission angles ($E_\gamma = 250$~MeV).
The red solid, black dotted, blue dashed, and magenta dash-dotted curves
 represent $\theta_\pi$=0$^\circ$, 15$^\circ$, 30$^\circ$, and 45$^\circ$, respectively.
 }
\label{fig:xs-250-angle}
\end{figure}
We examine the pion emission angle dependence of the $\gamma d\to \pi^+ nn$ cross
section as well.
The $M_{nn}$ distributions of different angles are shown in Fig.~\ref{fig:xs-250-angle}.
The emission angle is changed from $0^\circ$ to $45^\circ$ in the laboratory
frame.
In the small $M_{nn}$ region, which is the most sensitive to the $nn$
scattering length, the cross section is significantly larger for smaller
emission angles.
Thus, from a statistical viewpoint,
a small emission angle $\theta_\pi\sim 0^\circ$ is favorable
to measure the cross sections
and experimentally determine the $a_{nn}$.
As will be discussed in Sec.~\ref{sec:virtual},
$\theta_\pi = 0^\circ$ is also useful
for extracting photoproduction cross sections 
at the required precision and resolution
in an electron scattering experiment.
Thus, we use $\theta_\pi= 0^\circ$ in the calculations below.

\begin{figure}[t]
\begin{center}
\includegraphics[width=1\textwidth]{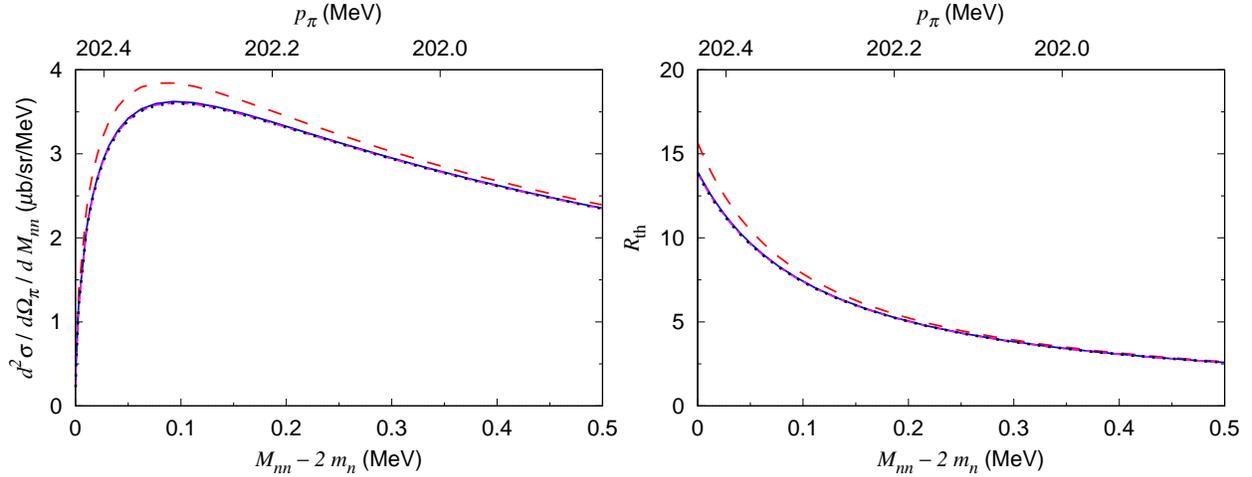}
\end{center}
 \caption{
(Left) The $M_{nn}$ distribution at $E_\gamma = 250$~MeV
 calculated with different $NN$ potentials.
 The red long-dashed, black dotted, blue solid, and magenta short-dashed
 curves are calculated with
 the CD-Bonn ($a_{nn} = -18.9$~fm), Reid93 ($a_{nn} = -17.3$~fm),
Nijmegen I ($a_{nn} = -17.3$~fm), and Nijmegen II ($a_{nn} = -17.3$~fm)
 $NN$ potentials, respectively.
The other conditions are the same as those in Fig.~\ref{fig:xs}.
 (Right)  Ratio $R_{\rm th}$ defined in Eq.~(\ref{eq:rth})
for the same condition as the left panel.
 }
\label{fig:xs-nn}
\end{figure}
We then examine various model dependences in
$d^2\sigma/d\Omega_\pi/dM_{nn}$ and $R_{\rm th}$ for $\gamma d\to\pi^+ nn$.
As we later observe, 
the shape of $R_{\rm th}$ in $M_{nn}-2m_n<0.5$~MeV 
(2~MeV $<M_{nn}-2m_n<6$~MeV)
is useful in determining the $a_{nn}$ ($r_{nn}$).
Thus, we give special attention to the
theoretical uncertainties on $R_{\rm th}$
and quantify it as follows:
\begin{eqnarray}
\Delta R_{\rm th} (M_{nn}) 
=  
\sqrt{
{1\over N_{\rm model}}\sum_{i=1}^{N_{\rm model}}
\left[
R^i_{\rm th} (M_{nn}) 
-R_{\rm th} (M_{nn})
\right]^2 
}
\ ,
\label{eq:err}
\end{eqnarray}
where $R_{\rm th} (M_{nn})$ is calculated with the standard setting 
used in Figs.~\ref{fig:xs}--\ref{fig:xs-250-angle},
while some dynamical content has been replaced by a different model in 
$R^i_{\rm th} (M_{nn})$.

\begin{figure}[t]
\begin{center}
\includegraphics[width=1\textwidth]{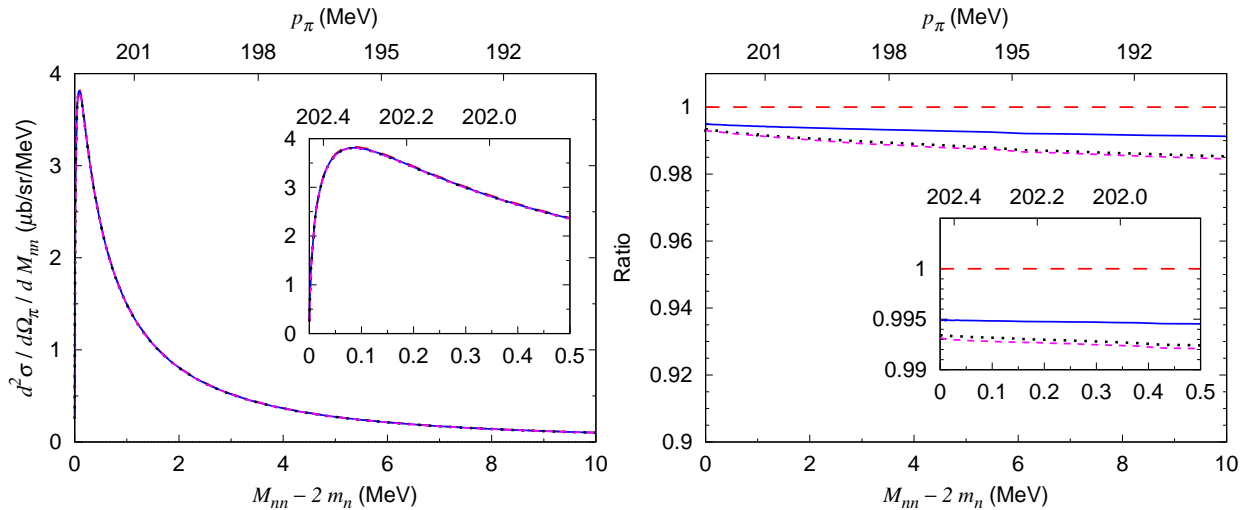}
\end{center}
 \caption{
 (Left) The $M_{nn}$ distribution at $E_\gamma = 250$~MeV calculated with 
the replacement in Eq.~(\ref{eq:amp_NN2}); $a_{nn} = -18.9$~fm, $r_{nn} = 2.75$~fm.
 The red long-dashed, black dotted, red solid, and magenta short-dashed
 curves are obtained with 
 the CD-Bonn, Reid93, Nijmegen I, and Nijmegen II potentials, respectively.
 The four curves are nearly the same.
 The other conditions are the same as those in Fig.~\ref{fig:xs}.
 (Right) Ratios of the curves in the left panel. 
Each of the curves in the left panel is divided by 
the one from the CD-Bonn potential and is shown with the same feature. 
 }
\label{fig:xs-nn-250-os}
\end{figure}
\begin{figure}[t]
\begin{center}
\includegraphics[width=1\textwidth]{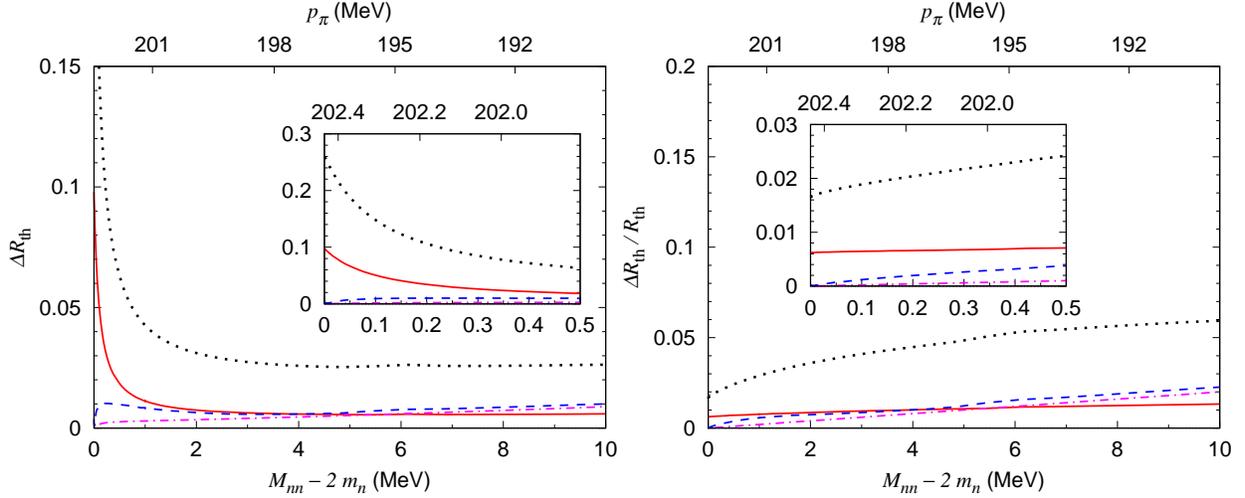}
\end{center}
 \caption{
Theoretical errors of $R_{\rm th}(M_{nn})$ defined in Eq.~(\ref{eq:err}).
(Left) The red solid, blue dashed, 
black dotted, and magenta dash-dotted curves
represent the uncertainties of the $NN$ potential, 
on-shell $\gamma p\to\pi^+ n$ amplitudes,
off-shell effects of $\gamma p\to\pi^+ n$ amplitudes,
and meson-exchange current effects, respectively.
(Right) Each curve in the left panel is divided by 
$R_{\rm th}(M_{nn})$ with the standard setting.
 }
\label{fig:error}
\end{figure}
\begin{figure}[t]
\begin{center}
\includegraphics[width=1\textwidth]{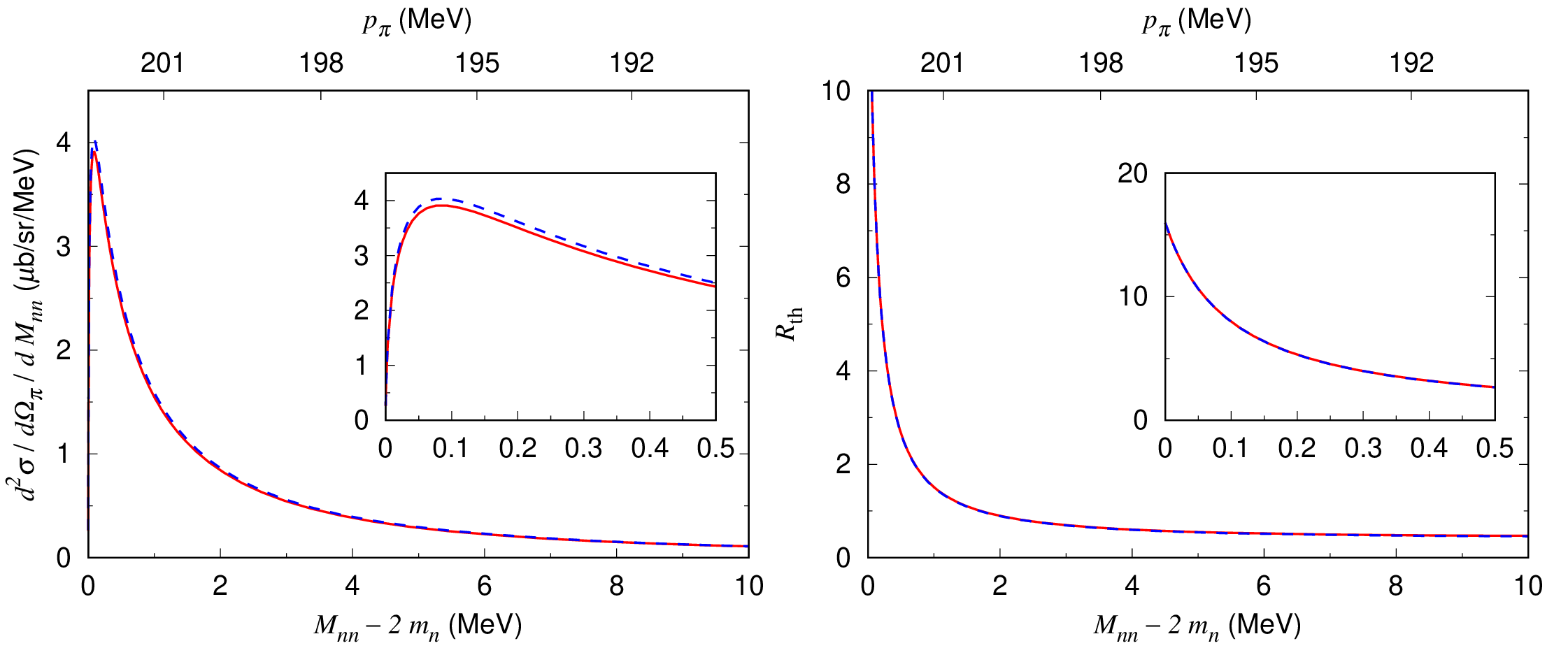}
\end{center}
 \caption{
(Left) The $M_{nn}$ distributions for different elementary amplitudes at $E_\gamma = 250$~MeV.
 The red solid and blue dashed curves are obtained using the
on-shell $\gamma p\to \pi^+ n$ amplitudes
from the DCC model and the CM12 parametrization~\cite{cm12},
respectively.
The other conditions are the same as those in Fig.~\ref{fig:xs}.
 (Right) Ratio $R_{\rm th}$ defined in Eq.~(\ref{eq:rth})
 for the same condition as the left panel.
}
\label{fig:xs-said}
\end{figure}

We first examine the $NN$ interaction model dependence.
We use four high-precision phenomenological $NN$ potentials to describe
the $NN$ rescattering and the deuteron wave function: the CD-Bonn potential~\cite{cdbonn},
the Nijmegen I potential~\cite{nij},
the Nijmegen II potential~\cite{nij},
and the Reid93 potential~\cite{nij}, 
for which the scattering length is 
$a_{nn} = -18.9, -17.3, -17.3$, and $-17.3$~fm, respectively,
and the effective range is
$r_{nn} = 2.8$~fm for all.
A comparison of the $d^2\sigma/d\Omega_\pi/dM_{nn}$
and $R_{\rm th}$ values calculated with these different $NN$
potentials is shown in Fig.~\ref{fig:xs-nn}.
We can see that 
the cross section with the CD-Bonn potential is 
$\sim$10\% larger than the others
around the peak region
due to the larger scattering length.
However, the cross sections are essentially the same among 
the Nijmegen I, II, and Reid93 potentials, for which
$a_{nn}$ and $r_{nn}$ are the same.
This result is non-trivial 
because
the $NN$ amplitudes from these three $NN$ models
have different off-shell behaviors.
In a previous study on $\pi^- d\to \gamma nn$~\cite{anders},
the authors found that
the neutron time-of-flight spectrum obtained with
the Nijmegen I potential
was non-negligibly different from that with
the Nijmegen II potential.

We further study the effects of the off-shellness of the $NN$
amplitude.
As observed in previous works on $\pi^-d\to\gamma nn$~\cite{GGS,anders},
the uncertainty of such off-shell behavior was the largest
source of the model dependence of the extracted $a_{nn}$.
We examine the $NN$ model dependence of the off-shell effect.
To clarify this, we adjust the on-shell amplitudes
of the different $NN$ models to be the same via using the following
replacement in Eq.~(\ref{eq:amp_NN}):
\begin{eqnarray}
 t_{NN}(M_{N_1N_2}) \Rightarrow  
t^{\rm ERE}_{NN,NN}(M_{N_1N_2}) 
\times
  \left[ {
t_{NN}(M_{N_1N_2}) 
\over  t^{\rm on\mbox{-}shell}_{NN,NN}(M_{N_1N_2})} \right] \ ,
\label{eq:amp_NN2}
\end{eqnarray}
where $t^{\rm ERE}_{NN,NN}$ is the $NN$ amplitude parametrized with the effective
range expansion of Eq.~(\ref{eq:ERE})
without the $O(p^4)$ contribution.
The result obtained with this replacement is shown in 
Fig.~\ref{fig:xs-nn-250-os}(left).
Within the considered high-precision $NN$ potentials,
the off-shell effect is very similar.
To clarify the model dependence, 
the ratios of the curves in Fig.~\ref{fig:xs-nn-250-os}(left)
divided by the one from the CD-Bonn potential
are shown in Fig.~\ref{fig:xs-nn-250-os}(right).
We also calculate
the $NN$ off-shell uncertainty of $R_{\rm th}$
using Eq.~(\ref{eq:err}) with $N_{\rm model} = 3$ for the 
Nijmegen I, II, and Reid93 potentials.
The calculated $\Delta R_{\rm th}$ is represented by the red solid curve in 
Fig.~\ref{fig:error}(left).
The variation of the shape of $R_{\rm th}$ due to the uncertainty could be
seen more clearly in $\Delta R_{\rm th}/R_{\rm th}$, as shown in 
Fig.~\ref{fig:error}(right).
From the figure, we can see that the uncertainty of the $NN$ off-shell
effect on $R_{\rm th}$ is less than 1.5\% (1\%) for 
$M_{nn}-2m_n \le 10$~MeV (0.5~MeV).

The uncertainty of the elementary $\gamma p\to\pi^+ n$ amplitudes may 
affect the theoretically calculated $M_{nn}$ distribution.
This uncertainty
could be examined via comparing calculations with different amplitude models.
We use the same DCC model as before and also
the Chew-Mandelstam (CM12) parametrization~\cite{cm12}.
The CM12 parametrization is a $K$-matrix fit to the $\gamma N\to\pi N$
data and, by construction, has only on-shell amplitudes.
Thus, for comparison, we also use the on-shell DCC amplitudes.
As shown in Fig.~\ref{fig:xs-said}(left),
the $M_{nn}$ distributions of $\gamma d\to\pi^+ nn$ at $E_\gamma = 250$~MeV
are calculated with the DCC and CM12 elementary on-shell amplitudes. 
In this particular kinematics ($\theta_\pi = 0^\circ$ and $M_{nn}\sim 2m_n$),
the cross sections obtained with the DCC model are slightly smaller 
than those obtained with the CM12 solution.
This reasonable agreement occurs because both the DCC model and
 CM12 parametrization well reproduce the $\gamma p\to\pi^+ n$ data.
Furthermore, the slight difference seen in Fig.~\ref{fig:xs-said}(left)
is almost perfectly removed in the
ratio $R_{\rm th}$, as shown in Fig.~\ref{fig:xs-said}(right).
This demonstrates that by analyzing $R_{\rm th}$ and its experimental counterpart, 
we can study low-energy $nn$ scattering
without being influenced by the uncertainty associated with 
on-shell elementary $\gamma p\to\pi^+ n$ amplitudes.
The uncertainty of $R_{\rm th}$
due to the on-shell $\gamma p\to\pi^+ n$ amplitudes
is calculated using Eq.~(\ref{eq:err}) with $N_{\rm model} = 1$ for the 
CM12, as shown in 
Fig.~\ref{fig:error} by the blue dashed curve.

\begin{figure}[t]
\begin{center}
\includegraphics[width=1\textwidth]{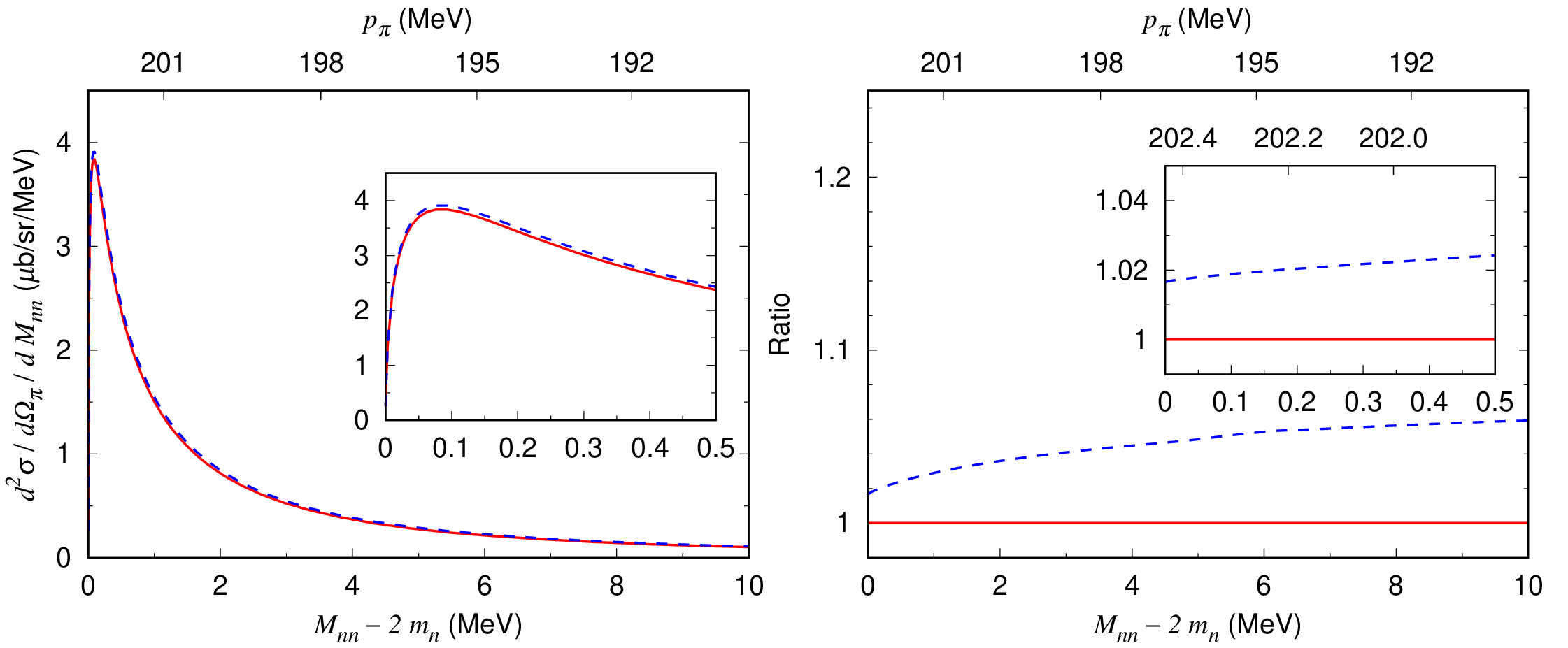}
\end{center}
 \caption{
(Left) 
The $M_{nn}$ distributions at $E_\gamma = 250$~MeV with and without 
the off-shell effects from the $\gamma p\to\pi^+ n$ amplitudes.
The red solid curve represents the original calculation, and 
the blue dashed curve is obtained by replacing 
the off-shell amplitudes with on-shell ones.
(Right) Ratios of the curves in the left panel.
Each of the curves is divided by the red solid curve in the left panel and is
 shown with the same feature. 
 }
\label{fig:xs-nn-250-dcc-os}
\end{figure}
In the loop diagrams,
the elementary $\gamma p\to \pi^+ n$ amplitudes also induce
uncertainty associated with their off-shell effects.
Although the off-shell behavior of the DCC model
has been constrained by fitting data to some extent, 
some uncertainty would still exist. 
Thus, we study the off-shell effect by using the on-shell elementary amplitudes in 
Eqs.~(\ref{eq:amp_NN}) and (\ref{eq:amp_MN})
and comparing the $R_{\rm th}$ from this calculation with 
the original value that considers the off-shell effect. 
The result is shown in Fig.~\ref{fig:xs-nn-250-dcc-os}.
We observe that the off-shell effect reduces the 
$R_{\rm th}$ and, thus, the cross sections by
1.7\%--2.4\% in $M_{nn}-2m_n \le 0.5$~MeV
and 4.0\%--6.0\% in 2~MeV $< M_{nn}-2m_n \le 6$~MeV.
The uncertainty of $R_{\rm th}$
is difficult to estimate because 
an off-shell $\gamma p\to \pi^+ n$ amplitude from a different model 
is not available; hence, we cannot study the model dependence.
Therefore, we make a conservative estimate of the uncertainty
due to the off-shell effects 
using Eq.~(\ref{eq:err}) with $N_{\rm model} = 1$ for the 
calculation with the on-shell DCC $\gamma p\to \pi^+ n$ amplitudes.
The result is shown
in Fig.~\ref{fig:error} by the black dotted curve.

Another source of theoretical uncertainty is the contributions from
meson-exchange currents aside from those included in Fig.~\ref{fig:diag}(c)
and not considered in the present model. 
Previous research on near-threshold $\gamma d\to \pi^+ nn$ 
based on the chiral perturbation theory~\cite{prev}
also did not consider such mechanisms because
they are higher-order effects 
within their counting scheme.
As we deal with the reactions at significantly higher photon energies, 
their argument does not necessarily apply to our case.
A previous chiral perturbation theory calculation 
of $\pi^-d \to nn\gamma$~\cite{anders}
considered more meson-exchange currents.
However, it was found that meson-exchange currents that can be accommodated
by Fig.~\ref{fig:diag}(c) provide a leading effect.
Therefore, we expect that the meson-exchange current missing in our calculation
provides a few \% contributions to the cross sections 
but the change in the shape would be even smaller.
Thus, we assume that the missing meson-exchange current linearly increase the 
$R_{\rm th} (M_{nn})$ of the standard setting as
$R^1_{\rm th} (M_{nn}) =
(1 + 0.002\, (M_{nn}-2m_n)/{\rm MeV})\times R_{\rm th} (M_{nn})$.
We then estimate the uncertainty of
$\Delta R_{\rm th}$ using Eq.~(\ref{eq:err}), as shown in 
Fig.~\ref{fig:error} by the magenta dash-dotted curve.

Let us summarize the various theoretical uncertainties shown in 
Fig.~\ref{fig:error}.
The largest uncertainty is from the off-shell effects 
of the elementary $\gamma p\to\pi^+n$ amplitudes
shown by the black dotted curve,
and it can change $R_{\rm th}$ by $\ltap 6$\%.
The other uncertainties are mostly $\ltap 2$\% effects on 
$R_{\rm th}$.
All the uncertainties are smaller in $M_{nn}-2m_n\le 0.5$~MeV where
$R_{\rm th}$ is sensitive to $a_{nn}$.
Furthermore, the uncertainties of the line shape of $R_{\rm th}$ are generally smaller
than the absolute value of $R_{\rm th}$.

\begin{figure}[t]
\begin{center}
\includegraphics[width=1\textwidth]{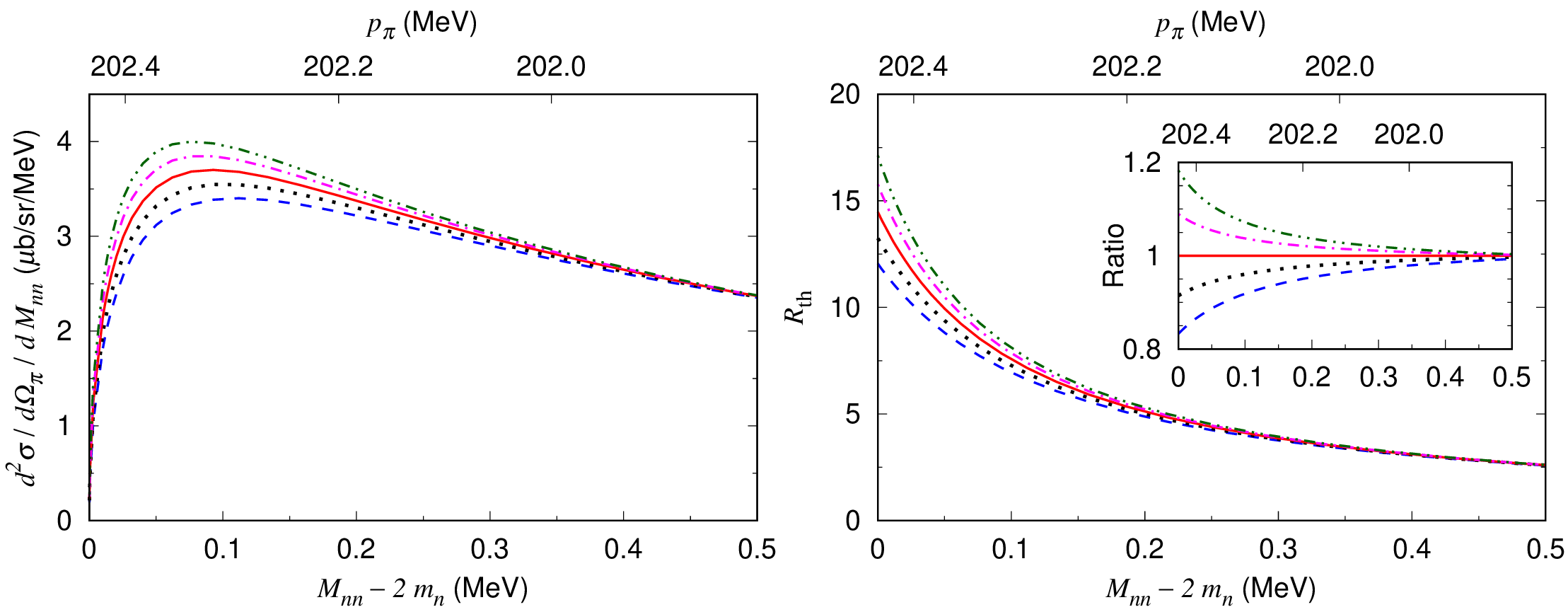}
\end{center}
 \caption{
The $M_{nn}$  distributions for different scattering length ($a_{nn}$) values.
 The blue dashed, black dotted, red solid,
 magenta dash-dotted, and green dash-two-dotted curves are calculated
 with $a_{nn} = -16, -17, -18, -19$, and $-20$~fm, respectively; $r_{nn} = 2.75$~fm.
 The off-shell dependence of the $NN$ amplitudes is drawn from the
 CD-Bonn potential, as explained in the text.
In the inlet, each $R_{\rm th}$ value is divided by 
the $R_{\rm th}$ of $a_{nn} = -18$~fm.
 The other conditions are the same as those in Fig.~\ref{fig:xs}.
 }
\label{fig:xs-nn-250-a}
\end{figure}
\begin{figure}[t]
\begin{center}
\includegraphics[width=1\textwidth]{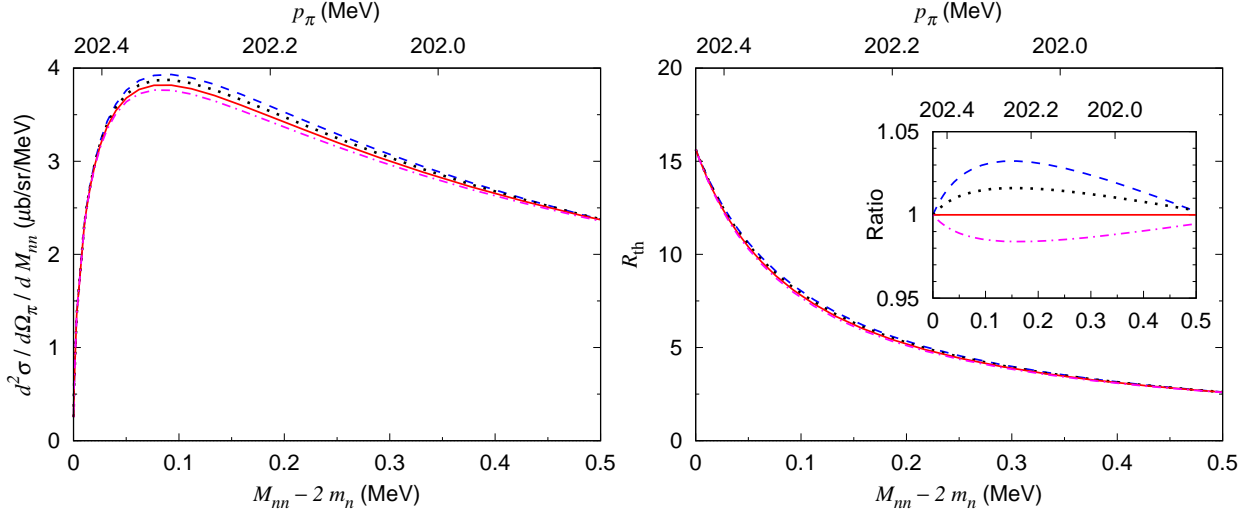}
\end{center}
 \caption{
The $M_{nn}$ distributions for different effective range ($r_{nn}$) values.
 The blue dashed, black dotted, red solid, and
 magenta dash-dotted curves are calculated
 with $r_{nn} = 1$, 2, 3, and 4~fm, respectively; $a_{nn} = -18.9$~fm.
In the inlet, each
$R_{\rm th}$ value is divided by 
the $R_{\rm th}$ of $r_{nn} = 3$~fm.
 The other conditions are the same as those in Fig.~\ref{fig:xs-nn-250-a}.
 }
\label{fig:xs-nn-250-r}
\end{figure}
\begin{figure}[b]
\begin{center}
\includegraphics[width=1\textwidth]{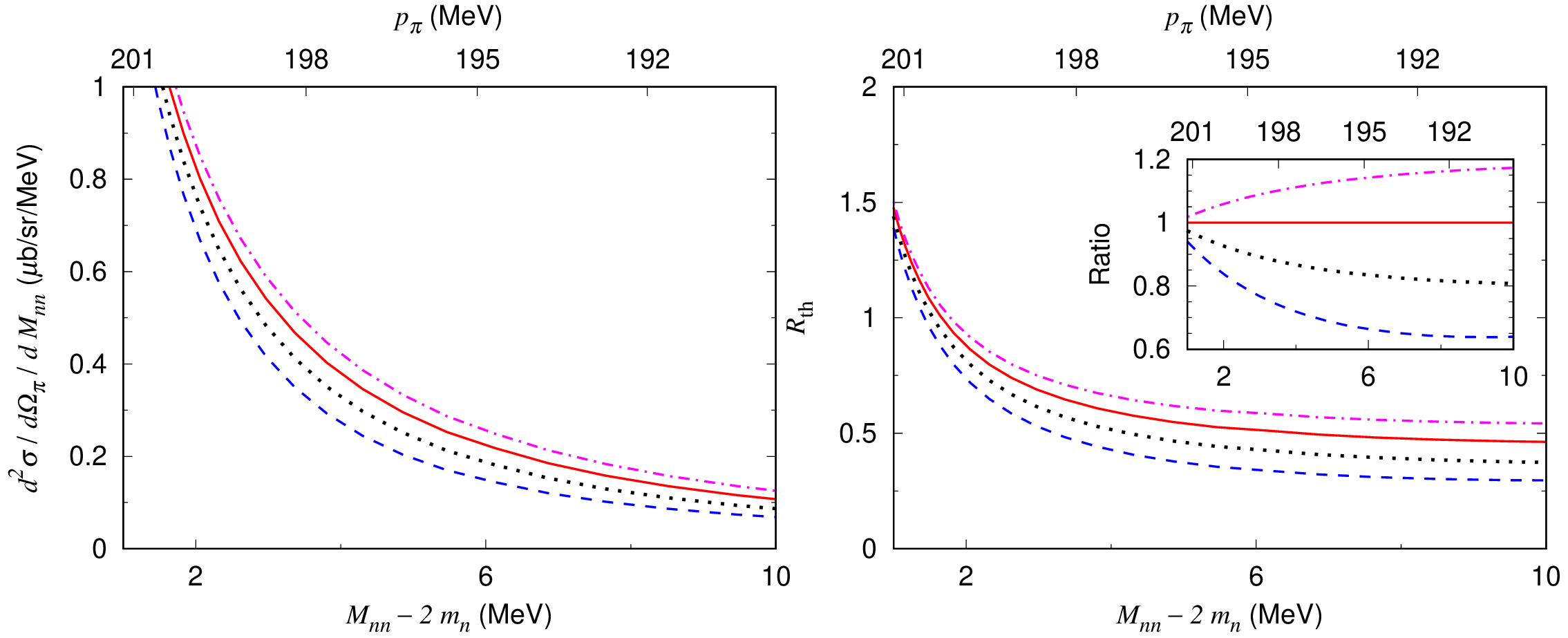}
\end{center}
 \caption{
 Same as in Fig.~\ref{fig:xs-nn-250-r} but for
1~MeV $\le M_{nn}-2m_n\le$ 10~MeV.
 }
\label{fig:xs-nn-250-r2}
\end{figure}
We now study the dependence of $d^2\sigma/d\Omega_\pi/dM_{nn}$
and $R_{\rm th}$ on scattering parameters $a_{nn}$ and $r_{nn}$.
For this purpose, 
we again use Eq.~(\ref{eq:amp_NN2}) to calculate the $NN$ rescattering
amplitudes. 
As discussed above, various model dependences change
the absolute magnitude of $d^2\sigma/d\Omega_\pi/dM_{nn}$
and $R_{\rm th}$ by a few \% levels, while their shapes
remain rather stable. 
To extract the
$a_{nn}$ and $r_{nn}$
at the precision of $\ltap 0.5$~fm,
which is comparable to previous determinations~\cite{huhn,gonz,wits,chen},
we use only the shape of 
($d^2\sigma/d\Omega_\pi/dM_{nn}$ and )
$R_{\rm th}$
to determine the scattering parameters. 
To start, we vary the $a_{nn}$ from $-16$~fm to $-20$~fm, with $r_{nn} = 2.75$~fm
fixed; the obtained 
$d^2\sigma/d\Omega_\pi/dM_{nn}$ and $R_{\rm th}$ are shown in 
Fig.~\ref{fig:xs-nn-250-a}.
The effect of changing the $a_{nn}$ value is only seen in a small region of
$M_{nn}-2m_n\ltap 0.3$~MeV.
When the $a_{nn}$ is changed by 1~fm,
the cross section changes by $\sim 10$\% at most at $M_{nn}\sim 2m_n$,
and by 4\%--5\% ($\sim$ 0.17~$\mu$b/sr/MeV)
at $M_{nn}-2m_n = 0.07$--0.10~MeV where the cross section reaches its peak.
As the shape of $d^2\sigma/d\Omega_\pi/dM_{nn}$
and, thus, $R_{\rm th}$ sensitively
changes as the $a_{nn}$ value changes,
fitting $R_{\rm th}$ to the corresponding data over 
$M_{nn}-2m_n\ltap 0.3$~MeV is an efficient way to precisely determine the $a_{nn}$.

Next, we study the $r_{nn}$ dependence.
Here, we vary the $r_{nn}$ from 1~fm to 4~fm, with fixed
$a_{nn} = -18.9$~fm.
The result for $M_{nn}-2m_n< 0.5$~MeV is given in Fig.~\ref{fig:xs-nn-250-r},
while that for 1~MeV $<M_{nn}-2m_n< 10$~MeV is given in Fig.~\ref{fig:xs-nn-250-r2}.
For the negative $a_{nn}$,
the $NN$ rescattering amplitude of Fig.~\ref{fig:diag}(b)
becomes weaker as the positive $r_{nn}$ increases.
Thus, as the $r_{nn}$ increases, 
the cross section reduces (increases)
in the small (large) $M_{nn}$ regions wherein 
the $NN$ rescattering (impulse) amplitude dominates.
When the $r_{nn}$ increases by 1~fm
at $M_{nn}-2m_n\sim 0.1$~MeV,
the cross section 
reduces by $\sim 1.5$\% ($\sim$ 0.06~$\mu$b/sr/MeV).
As discussed in the above paragraph, 
the shape of $R_{\rm th}$ is sensitive to the $a_{nn}$ 
in $M_{nn}-2m_n\ltap 0.3$~MeV.
In this $M_{nn}$ region, 
the shape of $R_{\rm th}$ also depends on the $r_{nn}$,
as seen in the inlet of Fig.~\ref{fig:xs-nn-250-r}.
If the $r_{nn}$ is in the range of 
$|r_{nn}-2.75\, {\rm fm}|< 0.5$~fm, as expected from
Eq.~(\ref{eq:ann}),
the shape of $R_{\rm th}$ changes from that for $r_{nn}$ = 2.75~fm
by less than 1\%.
Thus, it is important to control the $r_{nn}$ at the 0.5~fm level.
Although old data were used to constrain the
$r_{nn}$ (see Ref.~\cite{review_slaus} for review),
it would be more desirable to use the $\gamma d\to \pi^+nn$ data
to directly constrain the $r_{nn}$
at the precision of 0.5~fm,
without resorting to CS.
As seen in the inlet of Fig.~\ref{fig:xs-nn-250-r2}(right),
the shape of $R_{\rm th}$ for 2~MeV $<M_{nn}-2m_n< 6$~MeV 
shows a good sensitivity to the $r_{nn}$ but no sensitivity to the $a_{nn}$.

We then perform a Monte Carlo simulation and 
extract the $a_{nn}$ and $r_{nn}$ with the uncertainties from
$R_{\rm exp}$,
an experimental counterpart to $R_{\rm th}$,
under several realistic settings, such as different finite $M_{nn}$ bin widths and 
different precisions of $R_{\rm exp}$.
The $R_{\rm exp}$ data are generated with the 
fixed values of $a_{nn}^\circ$ and $r_{nn}^\circ$.
Specifically, we perform
a procedure defined by the following four steps:
\begin{enumerate}
 \item[(i)] We introduce:
$R^\circ_{\rm exp}(a_{nn}^\circ,r_{nn}^\circ;M_{nn})\equiv R_{\rm th}(a_{nn}^\circ,r_{nn}^\circ;M_{nn}) + g\Delta R_{\rm th}^{\rm all}(M_{nn})$
where $\Delta R_{\rm th}^{\rm all}(M_{nn})$ is 
the quadratic sum of 
$\Delta R_{\rm th} (M_{nn})$s from different sources, as shown in Fig.~\ref{fig:error}(left).
The parameter $g$ is randomly generated at each cycle 
according to the standard normal distribution.
 \item[(ii)]
The histogram $R^\circ_{\rm exp}(a_{nn}^\circ,r_{nn}^\circ; i)$ for the $i$-th bin is
created by averaging $R^\circ_{\rm exp}(a_{nn}^\circ,r_{nn}^\circ; M_{nn})$
with respect to $M_{nn}$ over the bin width.
 \item[(iii)]
The histogram data $R_{\rm exp} (i)$ are generated from $R^\circ_{\rm exp}(i)$
via including a statistical fluctuation corresponding to the given precision.
 \item[(iv)]
We simultaneously search for the $a_{nn}$ and $r_{nn}$ 
with which $R_{\rm exp}(a_{nn}, r_{nn}; M_{nn})$, averaged
	    over the bin width, optimally fits
$R_{\rm exp}(i)$ in the $M_{nn}$ range of [0.00, 6.00) MeV. 
We can multiply a free overall coefficient
to $R_{\rm exp}(a_{nn}, r_{nn}; M_{nn})$,
as we just attempt to reproduce the shape.
\end{enumerate}
The above procedure is repeated 10,000 times, and 
the width of the obtained $a_{nn}$ ($r_{nn}$) distribution corresponds to the uncertainty.
This analysis is performed with several different values of 
$a_{nn}^\circ$ and $r_{nn}^\circ$ over
$-20$~fm $\le a_{nn}^\circ\le -15$~fm and 1~fm $\le r_{nn}^\circ\le 5$~fm.
The centroid values of the $a_{nn}$ and $r_{nn}$ 
well reproduce $a_{nn}^\circ$ and $r_{nn}^\circ$.
The $a_{nn}$ uncertainty is approximately proportional to the $R_{\rm exp}$ precision.
When the $M_{nn}$ bin width is 0.04~MeV, 
a precision of 5\% is required to lower the $a_{nn}$ uncertainty to less than 0.5~fm.
With the same bin width, 
the $r_{nn}$ uncertainty is $\ltap 0.1$~fm ($\sim$ 0.05~fm)
for the precision of 5\% (0\%).
A smaller bin width results in a smaller $a_{nn}$ uncertainty
($\Delta a_{nn} = 0.13$--0.27~fm for the bin width of 0.01--0.08~MeV 
and 2\% precision)
but a larger $r_{nn}$ uncertainty
($\Delta r_{nn} = 0.23$--0.06~fm for the bin width of 0.01--0.08~MeV
and 2\% precision)
owing to the theoretical uncertainties of $\Delta R_{\rm th}$s.
These results do not change very much within the specified range of 
$a_{nn}^\circ$.
When we use $a_{nn}^\circ = -18.9$~fm and $r_{nn}^\circ = 2.75$~fm,
we obtain $\Delta a_{nn}=0.21$~fm and $\Delta r_{nn}=0.06$~fm
for an 0.04~MeV bin width 
and $\pm 2\%$ precision of $R_{\rm exp}$ in each bin.
By comparing the results obtained with and without 
$\Delta R_{\rm th}^{\rm all}$, 
we find that 
$\Delta a_{nn}$(theory)=0.03~fm and $\Delta r_{nn}$(theory)=0.06~fm
due to $\Delta R_{\rm th}^{\rm all}$
contribute to $\Delta a_{nn}$ and $\Delta r_{nn}$ 
through the quadratic sum.

\section{Experimental strategy used to extract high-resolution 
$\gamma d\to\pi^+ nn$ data from pion electroproduction data
}
\label{sec:virtual}

In this section, 
we discuss how we experimentally obtain
high-resolution $M_{nn}$ distribution data of $\gamma d\to\pi^+ nn$ 
to determine the $nn$ scattering parameters. 
Since we do not detect neutrons,
the momenta and angles of the photon and pion have to be measured with
sufficiently high resolutions.
Considering the currently available experimental facilities around the world,
we cannot achieve such high resolutions with a real photon beam.
However, we can achieve a high $M_{nn}$ resolution utilizing
virtual photons ($\gamma^*$s) from electron scattering 
and two magnetic spectrometers to detect the scattered electrons and emitted positive pions.
Upon tuning the electron scattering kinematics, we can measure cross sections for
pion production with a so-called ``almost-real'' photon
at a low momentum transfer ($Q^2\sim 0$).

The triple-differential unpolarized cross section for the $d(e,e'\pi^+)nn$ reaction is
written as follows (differentiation with respect to
$M_{nn}$ being omitted):
\begin{equation}
\displaystyle
\frac{d^3\sigma^{e\,d}}{dE_{e'}\, d\Omega_{e'}\, d\Omega_{\pi}}
=  \Gamma_\gamma \left(
\frac{d\sigma_{\rm T}}{d\Omega_\pi}+
\epsilon_{\rm L} \frac{d\sigma_{\rm L}}{d\Omega_\pi}+
\sqrt{2\epsilon_{\rm L}\left(1+\epsilon\right)} \frac{d\sigma_{\rm LT}}{d\Omega_\pi}
\cos \phi_\pi+
\epsilon \frac{d\sigma_{\rm TT}}{d\Omega_\pi}\cos 2\phi_\pi
\right),
\label{eq:es}
\end{equation}
where the electron mass is neglected, and 
$\sigma_{\rm T}$,  $\sigma_{\rm L}$,  $\sigma_{\rm LT}$, and
$\sigma_{\rm TT}$
are the transverse, longitudinal, 
longitudinal-transverse interference,
and transverse-transverse interference cross sections
for $\gamma^*d\to\pi^+ nn$ in the laboratory frame, 
respectively~\cite{forest,mulders}.
The pion angle $\Omega_\pi=(\theta_\pi,\phi_\pi)$ is measured with respect to
the virtual photon direction, and 
$\phi_\pi$ is 
the angle between the electron-scattering $(e,e')$ plane and the pion-production 
$(\gamma^*,\pi)$ plane.
We denote the incident (scattered) electron energy and momentum in the
laboratory frame
by $E_e$ and $\bm{p}_e$ ($E_{e'}$ and $\bm{p}_{e'}$), respectively, 
and 
the electron scattering angle by 
$\cos \theta_{e'}=\left(\bm{p}_e\cdot \bm{p}_{e'}\right) /\left( |\bm{p}_e||\bm{p}_{e'}| \right)$.
Also,  
the four-momentum transfer from the electron to the deuteron is denoted
by $(\omega, \bm{q})=(E_e-E_{e'}, \bm{p}_e-\bm{p}_{e'})$
and the squared momentum transfer by
$Q^2=-\left(\omega^2-\left|\bm{q}\,\right|^2\right)$.
With these notations, we introduced in Eq.~(\ref{eq:es})
\begin{equation}
\epsilon=\left(1+\frac{2\left|\bm{q}\,\right|^2}{Q^2}\tan^2\frac{\theta_{e'}}{2}
    \right)^{-1} \quad {\rm and} \quad
\epsilon_{\rm L}=\frac{Q^2}{\omega^2} \epsilon\ ,
\end{equation}
and the virtual photon flux given by
\begin{equation}
\Gamma_\gamma=\frac{\alpha}{2\pi^2Q^2} 
\frac{E_\gamma}{1-\epsilon}\frac{E_{e'}}{E_e} \ ,
\end{equation}
with $\alpha$ as the fine structure constant, and 
$E_\gamma = \omega-Q^2/2m_d$ as
the photon equivalent energy in the laboratory frame
with which a real photon excites a deuteron to a hadronic system with
the same invariant mass as a virtual photon with the four-momentum $q$ does.
The virtual photon emission angle $\theta_\gamma$ is specified by
$\cos\theta_\gamma = \left(\bm{p}_e\cdot \bm{q}\right) /\left( |\bm{p}_e||\bm{q}| \right)$.
In Eq.~(\ref{eq:es}),
all cross sections depend on $E_\gamma$ and $Q^2$,
and $d\sigma_{\rm T}/d\Omega_\pi (Q^2 = 0)$ 
differentiated with respect to $M_{nn}$
corresponds to Eq.~(\ref{eq:dsigp-nn}).

The elementary $\gamma N\to\pi N$ data 
[$d\sigma^{\gamma N}_{\rm T}/d\Omega_\pi (Q^2 = 0)$]
are generally more precisely measured than
$\gamma^* N\to\pi N$ data [$d\sigma^{\gamma^* N}_{\rm T}/d\Omega_\pi (Q^2\ne 0)$]
from pion electroproductions.
A primary reason for this is that an uncertainty enters into the data when separating 
$d\sigma^{\gamma^* N}_{\rm T}/d\Omega_\pi$ from
 $d\sigma^{\gamma^* N}_{\rm L}/d\Omega_\pi$,
 $d\sigma^{\gamma^* N}_{\rm LT}/d\Omega_\pi$, and
 $d\sigma^{\gamma^* N}_{\rm TT}/d\Omega_\pi$.
Consequently, the $Q^2 = 0$ sector of the DCC model, as well as other similar models 
for $\gamma^{(*)} N\to\pi N$,
is significantly better tested by the real photon data,
in comparison with the finite $Q^2$ sector of the models.
Therefore, 
for a reliable determination of $nn$ scattering parameters
using such models,
almost-real photon data of $\gamma d\to\pi^+ nn$, i.e.,
$d\sigma_{\rm T}/d\Omega_\pi$ in Eq.~(\ref{eq:es}) at $Q^2\sim 0$,
 are highly preferred.

Here, we consider a method to extract 
$d\sigma_{\rm T}/d\Omega_\pi$ at $Q^2\sim 0$
from the $d(e,e'\pi^+)nn$ cross sections.
The $d\sigma_{\rm LT}/d\Omega_\pi$ and 
 $d\sigma_{\rm TT}/d\Omega_\pi$ terms in Eq.~(\ref{eq:es})
 vanish at $\theta_\pi=0$
because they are proportional to 
$\sin \theta_\pi$ and $\sin^2 \theta_\pi$, respectively.
Fortunately, the $\gamma d\to\pi^+nn$ data at $\theta_\pi = 0$ are exactly what
we need, as discussed in connection with Fig.~\ref{fig:xs-250-angle}.
Now,
Eq.~(\ref{eq:es}) at $\theta_\pi = 0$ is simplified:
\begin{equation}
\displaystyle
\frac{d^3\sigma^{e\,d}}{dE_{e'}\, d\Omega_{e'}\, d\Omega_{\pi}}
=  \Gamma_\gamma \left(
\frac{d\sigma_{\rm T}}{d\Omega_\pi}+
\epsilon_{\rm L} \frac{d\sigma_{\rm L}}{d\Omega_\pi}
\right).
\label{eq:es2}
\end{equation}
The $\epsilon_{\rm L} d\sigma_{\rm L}/d\Omega_\pi$ contribution can be made smaller
by using the kinematics wherein $\epsilon_{\rm L}$ and $Q^2$ 
are low.
Even when 
the $\epsilon_{\rm L} d\sigma_{\rm L}/d\Omega_\pi$ contribution cannot be made
negligible, we can still separate out ${d\sigma_{\rm T}}/{d\Omega_\pi}$ 
utilizing the linear $\epsilon_{\rm L}$ dependence in Eq.~(\ref{eq:es2}).

The A1 spectrometer facility at Mainz Microtron (MAMI)~\cite{spek} is an outstanding
candidate 
to conduct an experiment under the above-mentioned kinematic and
precision conditions.
The facility is capable of
providing high energy-resolution electron beams 
($\delta p/p<10^{-4}$) and
measuring
the momenta and angles of electrons and pions with the
high resolutions, 
i.e.,
$\delta p/p = 10^{-4}$ and $\delta\theta <3$~mrad ($0.2^\circ$),
required to determine the $nn$ scattering parameters.
Three magnetic spectrometers, i.e., SpekA, SpekB, and SpekC, are
placed in a horizontal plane ($\phi_\pi = 0^\circ$).
SpekA and SpekC, each of which covers $\pm 100$ mrad ($\pm 5.7^\circ$),
can be placed from $18^\circ$ to $160^\circ$ from the primary
electron beam direction. 
SpekB can be placed at more forward angles from $7^\circ$ to $62^\circ$
and has a relatively smaller coverage of $\pm 20$ mrad ($\pm 1.1^\circ$).
Table~\ref{tbl1} briefly lists the parameters of these three
spectrometers at MAMI.
\begin{table}[t]
\caption{Summary of the parameters of the three spectrometers at MAMI.
These parameters are taken from Ref.~\cite{spek}.
}\label{tbl1}
\begin{center}
\begin{tabular}{lrrr}
\hline
spectrometer & SpekA & SpekB & SpekC \\
\hline
minimum angle & $18^\circ$& $7^\circ$& $18^\circ$ \\
maximum angle & $160^\circ$& $62^\circ$& $160^\circ$ \\
\hline
horizontal angular coverage (mrad) & $\pm  100$ & $\pm  20$ & $\pm  100$ \\
vertical angular coverage (mrad) & $\pm  70$ & $\pm  70$ & $\pm  70$ \\
angular resolution at the target position (mrad) & $< 3$ & $< 3$ & $< 3$\\
\hline
momentum bite & $\pm 20\%$& $\pm 15\%$& $\pm 25\%$\\
momentum resolution ($\delta p/p$) & $10^{-4}$ & $10^{-4}$ & $10^{-4}$\\
\hline
\end{tabular}
\end{center}
\end{table}

\begin{figure}[t]
\begin{center}
\includegraphics[width=0.8\textwidth]{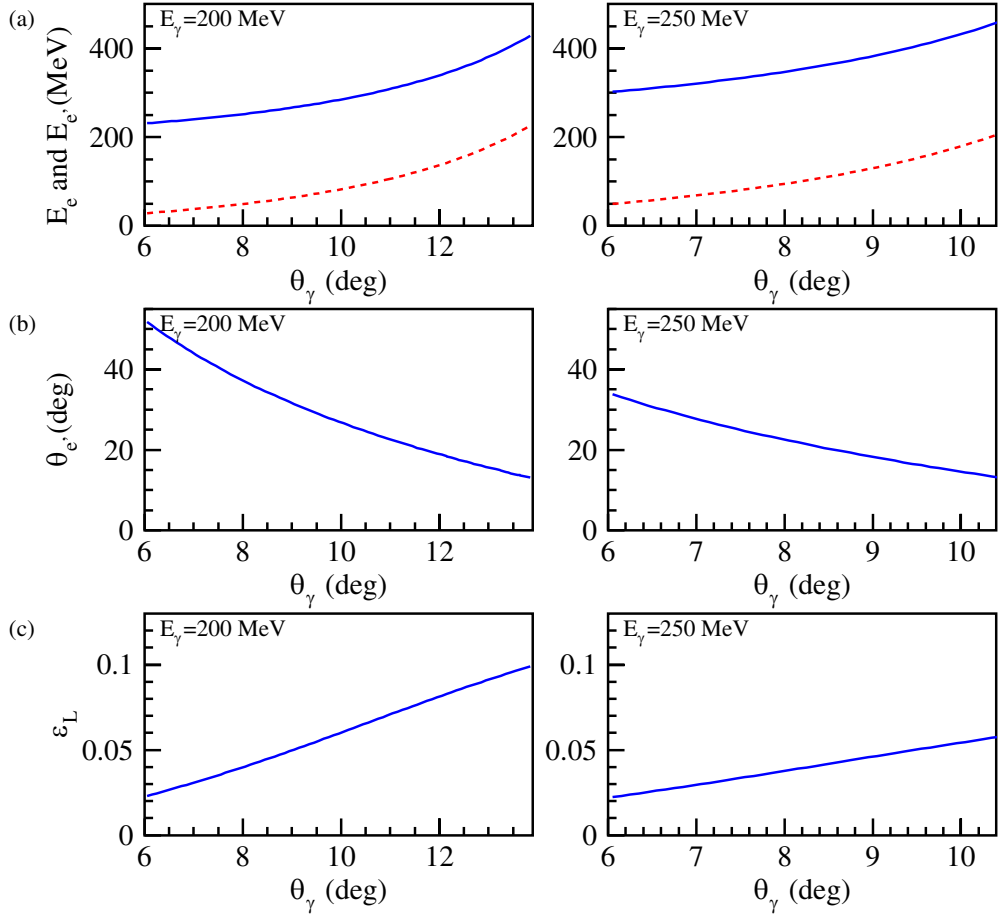}
\end{center}
\caption{(a) Incident electron energy $E_e$ (blue solid line) 
and scattered electron energy $E_{e'}$ (red dashed line)
as a function of the emission angle of virtual photons
in the laboratory frame $\theta_\gamma$ 
at $Q^2 = 0.0050$ GeV${}^2/c^2$ for $E_\gamma = 200$ MeV (left) and 250 MeV (right).
(b) Electron scattering angle $\theta_{e'}$.
(c) Degree of longitudinal polarization $\epsilon_{\rm L}$.
}\label{fig:spek}
\end{figure}

The experimental constraints do not allow us to use a kinematical
setting wherein
$\epsilon_{\rm L} d\sigma_{\rm L}/d\Omega_\pi$ is negligible
compared with $d\sigma_{\rm T}/d\Omega_\pi$.
Thus, a realistic solution is to separate 
the $d\sigma_{\rm T}/d\Omega_\pi$ and 
$d\sigma_{\rm L}/d\Omega_\pi$ contributions
by taking advantage of the linear $\epsilon_{\rm L}$ dependence 
in Eq.~(\ref{eq:es2}).
For this purpose, we need to measure cross sections
at several kinematical points
wherein
$\omega$ and $Q^2$,
thus, $d\sigma_{\rm T}/d\Omega_\pi$ and $d\sigma_{\rm L}/d\Omega_\pi$
are the same, while 
$\epsilon_{\rm L}$ is different.
In addition, the $Q^2$ has to be low enough 
to regard the virtual photon as real.
Suppose we utilize SpekA and SpekB to detect the scattered electrons
and emitted positive pions, respectively.
To obtain $E_\gamma = 200$ (250) MeV under the constraints of 
$\theta_{e'}\ge 12.3^\circ$ and $\theta_\gamma \ge 5.9^\circ$,
the minimum $Q^2$ achievable is 0.0021 (0.0013) GeV${}^2/c^2$
and $\epsilon_{\rm L} = 1.92\%$. 
At this $Q^2$ value, the experimental constraints would not allow us to
change $\epsilon_{\rm L}$; thus, the $\sigma_{\rm T}$-$\sigma_{\rm L}$ separation
is impossible.
To cover a wide range of $\epsilon_{\rm L}$ for the
separation, we choose $Q^2 = 0.0050$ GeV${}^2/c^2$.
To use the virtual photon of $Q^2 = 0.0050$ GeV${}^2/c^2$ and
$E_\gamma = 200$ or 250 MeV,
relations between the kinematical electron and photon variables
($E_e$, $E_{e'}$, $\theta_{e'}$, $\epsilon_{\rm L}$, and $\theta_\gamma$)
 are shown in Fig.~\ref{fig:spek}.
The experimental constraint of $\theta_\gamma\ge 5.9^\circ$
requires $E_e\ge 230$ (300) MeV
for $E_\gamma = 200$ (250)~MeV,
producing $\epsilon_{\rm L}\ge 2.24\%$ (2.17\%). 
The $d(e,e'\pi^+)nn$ cross sections 
measured at several electron kinematics in Fig.~\ref{fig:spek}
are used for the $\sigma_{\rm T}$-$\sigma_{\rm L}$ separation,
providing $d\sigma_{\rm T}/d\Omega_\pi$ at $Q^2 = 0.0050$ GeV${}^2/c^2$.

Note that within the $\gamma d\to\pi^+nn$ model used in this work,
$d\sigma_{\rm T}/d\Omega_\pi$ at $Q^2 = 0$ is slightly larger 
by $\sim 1$\% than that at $Q^2 = 0.0050$ GeV${}^2/c^2$, and 
the shape of the $M_{nn}$ distribution hardly changes.
Therefore, 
we can directly apply the analysis method and results
for the real photon
presented in the previous sections
to the data of 
$d\sigma_{\rm T}/d\Omega_\pi$ at $Q^2=0.0050$ GeV${}^2/c^2$
to determine the $nn$ scattering parameters.
Additionally, due to the high angle resolution ($<$ 3~mrad) of the facility, 
we do not need to modify the presented results 
in which the finite angle resolution is not accounted for.

\section{Summary}
\label{sec:summary}
In this paper, we discussed the possibility of extracting 
the low-energy neutron-neutron scattering parameters of
$a_{nn}$ and $r_{nn}$ from 
$\gamma d\to \pi^+nn$ cross-section data.
The analysis was based on a theoretical model of 
$\gamma d\to \pi^+nn$ that incorporated realistic 
elementary amplitudes for $\gamma p\to \pi^+n$, $NN\to NN$, 
and $\pi N\to \pi N$.
We demonstrated that $\gamma d\to \pi^+nn$ at 
the special kinematics with $\theta_\pi=0^\circ$
and $E_\gamma\sim 250$~MeV is suitable for studying 
neutron-neutron scattering in a low $nn$ invariant mass region ($M_{nn}\sim 2 m_n$)
 because the $NN$ rescattering mechanism dominates while 
the $\pi N\to \pi N$ rescattering contribution is negligible.
We assessed theoretical uncertainties
from various sources, including the
on- and off-shell behaviors of the $\gamma p\to \pi^+n$ and
$NN\to NN$ amplitudes.
This assessment showed that the shape of the ratio
$R_{\rm th}$, defined with the
$\gamma d\to \pi^+nn$ cross section ($d^2\sigma/dM_{nn}/d\Omega_\pi$)
as in Eq.~(\ref{eq:rth}),
was particularly useful for extracting the $a_{nn}$ and $r_{nn}$ from data
because it had
a good sensitivity to these scattering parameters and
 significantly reduced the model dependences compared with 
$d^2\sigma/dM_{nn}/d\Omega_\pi$.
The experimental counterpart to $R_{\rm th}$, $R_{\rm exp}$, 
is defined
 with measurable 
$\gamma d\to \pi^+nn$ and $\gamma p\to \pi^+n$ cross sections
and the deuteron wave function.
Through a Monte Carlo simulation,
we found that $R_{\rm exp}$ with 2\% error,
resolved into an $M_{nn}$ bin width of 0.04 MeV,
could determine $a_{nn}$ and $r_{nn}$ values with uncertainties of $\pm 0.21$ fm 
and $\pm 0.06$ fm, respectively,
if $a_{nn} = -18.9$~fm and $r_{nn} = 2.75$~fm.
The uncertainties did not significantly change when 
the $a_{nn}$ value was changed from $-20$ to $-15$~fm.
Such a high $M_{nn}$ resolution can be achieved with
an electron scattering experiment that utilizes the A1 spectrometer 
facility at MAMI.
The $d^2\sigma/dM_{nn}/d\Omega_\pi$ for $d(\gamma,\pi^+)nn$
can be separated from the
$d^4\sigma/dE_{e'}/d\Omega_{e'}/dM_{nn}/d\Omega_\pi$ 
for $d(e,e'\pi^+)nn$
at different $\epsilon_{\rm L}$ values but the same $Q^2\simeq 0$
 (an almost-real photon condition).
Since the proposed method does not require the difficult experimental task of
handling neutron detection efficiency and its uncertainty,
it
has a great advantage over the previous method
that extracted the $a_{nn}$ from the neutron time-of-flight spectrum of
$\pi^- d\to\gamma nn$.

\begin{acknowledgments}
This work was supported in part by the Japan Society for the Promotion of Science (JSPS) through Grants-in-Aid 
for Scientific Research (B) No.\ 19H01902, and
for Scientific Research on Innovative Areas Nos.\ 19H05104, 19H05141 and
 19H05181,
and also by 
National Natural Science Foundation of China (NSFC) under contracts 11625523.
\end{acknowledgments}

\end{document}